\pdfoutput=1
\documentclass[11pt,twoside,a4paper,cmspaper,final,collab]{cms-tdr}
\def\svnVersion{4f50501}\def\svnDate{2022/02/02}\def\cmsCernNoTag{CERN-EP-2021-093}\def\cmsCernDate{\today}\def\cmsMessage{Published in the European Physical Journal C as \href{http://dx.doi.org/10.1140/epjc/s10052-022-10127-0}{\doi{10.1140/epjc/s10052-022-10127-0}.}}
\begin{document}\cmsNoteHeader{HIG-19-007}

\newlength\cmsFigWidth
\newlength\cmsTabSkip\setlength{\cmsTabSkip}{1ex}
\setlength\cmsFigWidth{0.48\textwidth}
\ifthenelse{\boolean{cms@external}}{\providecommand{\cmsLeft}{top\xspace}}{\providecommand{\cmsLeft}{left\xspace}}
\ifthenelse{\boolean{cms@external}}{\providecommand{\cmsRight}{bottom\xspace}}{\providecommand{\cmsRight}{right\xspace}}

\newcommand{\ZOne}{\ensuremath{\PZ_1}\xspace}
\newcommand{\ZTwo}{\ensuremath{\PZ_2}\xspace}
\newcommand{\zd}{\ensuremath{\PZ_{\mathrm{D}}}\xspace}
\newcommand{\esp}{\ensuremath{\varepsilon}\xspace}
\newcommand{\ttNew}{\ensuremath{\PQt\PAQt}\xspace}
\newcommand{\ttV}{\ensuremath{\PQt\PAQt\PZ}\xspace}
\newcommand{\zg}{\ensuremath{\PZ\PGg}\xspace}
\newcommand{\wz}{\ensuremath{\PW\PZ}\xspace}
\newcommand{\zj}{\ensuremath{\PZ+\text{jets}}\xspace}
\newcommand{\cZh}{\ensuremath{C_{\PZ\PH}}\xspace}
\newcommand{\cZhLambda}{\ensuremath{C_{\PZ\PH}/\Lambda}\xspace}
\newcommand{\cah}{\ensuremath{C_{\Pa\PH}}\xspace}
\newcommand{\cahLambda}{\ensuremath{C_{\Pa\PH}/\Lambda^2}\xspace}
\newcommand{\usedLumi}{\ensuremath{137\fbinv}\xspace}
\newcommand{\ZX}{\ensuremath{\PZ \PX}\xspace}
\newcommand{\XX}{\ensuremath{\PX \PX}\xspace}
\newcommand{\JHUGen}{\textsc{JHUGen}\xspace}
\newcommand{\fourMu}{\ensuremath{4\PGm}\xspace}
\newcommand{\fourEl}{\ensuremath{4\Pe}\xspace}
\newcommand{\twoElTwoMu}{\ensuremath{2\Pe2\PGm}\xspace}
\newcommand{\twoMuTwoEl}{\ensuremath{2\PGm2\Pe}\xspace}
\newcommand*{\mass}[1]{\ensuremath{m_{#1}}\xspace}

\cmsNoteHeader{HIG-19-007} 
\title{Search for low-mass dilepton resonances in Higgs boson decays to four-lepton final states in proton-proton collisions at \texorpdfstring{$\sqrt{s}=13\TeV$}{sqrt(s)=13 TeV}}
\titlerunning{Search for low-mass dilepton resonances in Higgs boson decays to 4$\ell$ final states in pp collisions at 13\TeV}
\abstract{
  A search for low-mass dilepton resonances in Higgs boson decays is conducted in the four-lepton final state. The decay is assumed to proceed via a pair of beyond the standard model particles, or one such particle and a \PZ boson. The search uses proton-proton collision data collected with the CMS detector at the CERN LHC, corresponding to an integrated luminosity of 137\fbinv, at a center-of-mass energy $\sqrt{s} = 13\TeV$. No significant deviation from the standard model expectation is observed. Upper limits at 95\% confidence level are set on model-independent Higgs boson decay branching fractions. Additionally, limits on dark photon and axion-like particle production, based on two specific models, are reported.
}

\hypersetup{%
pdfauthor={CMS Collaboration},%
pdftitle={Search for low-mass dilepton resonances in Higgs boson decays to four-lepton final states in proton-proton collisions at sqrt(s) = 13 TeV},
pdfsubject={CMS},%
pdfkeywords={CMS, dark photon, axion like particles, Higgs}}

\date{\today}
\maketitle 

\section{Introduction} 
\label{sec:intro}

Following the discovery of the Higgs boson (\PH) by the ATLAS and CMS Collaborations~\cite{paper:Aad:2012,paper:Chatrchyan:2012,Chatrchyan:2013lba} at the CERN LHC, a thorough program of precise measurements~\cite{CMS:2021ugl,Sirunyan:2021fpv,CMS:2021kom} has been carried out to uncover possible deviations from the standard model (SM) or to decipher the nature of the Higgs sector. In particular, various exotic decays of the Higgs boson have been considered, in which small deviations in the Higgs boson decay 
width or discovery of exotic decay modes could constitute evidence of beyond the SM (BSM) physics.

This paper describes a search for exotic decays of the Higgs boson ${\PH \to \PZ \PX}$ or ${\PH \to \PX \PX}$ in the four-lepton (electrons or muons) final state, using a sample of proton-proton collision data at a center-of-mass energy of 13\TeV recorded by the CMS experiment in 2016--2018. The analyzed data sample corresponds to an integrated luminosity of \usedLumi.
Here \PX represents a possible BSM particle that could decay into a pair of opposite-sign, same-flavor (OSSF) leptons. In this paper, we consider two specific BSM models. 
In both models, leptonic decays of \PX and \PZ to either two muons or electrons give rise to the 4$\ell$ (where $4\ell$ may denote $4\PGm$, $2\Pe2\PGm$, or $4\Pe$) final states. 
Assuming narrow-width approximation decays of \PX, only the mass range $m_\PX < m_\PH - m_\PZ \approx 35\GeV$ ($m_\PX < m_\PH/2 \approx 62.5\GeV$) 
is kinematically possible for $\PH \to \PZ \PX$ ($\PH \to \PX \PX$), where $\mass{\PH}$ and $\mass{\PZ}$ are the Higgs boson mass and Z boson mass, respectively. The decay channel $\Pp\Pp \to \PH \to 4\ell$ has a large signal-to-background ratio. This channel allows a complete reconstruction of the kinematics of the Higgs boson based on final-state decay particles.  
In this analysis, a mass range of $4.0 < m_{\PX} < 35.0\GeV$ (62.5\GeV) is considered.

\begin{figure*}[!htbp]
\centering
    \includegraphics[width=0.33\textwidth]{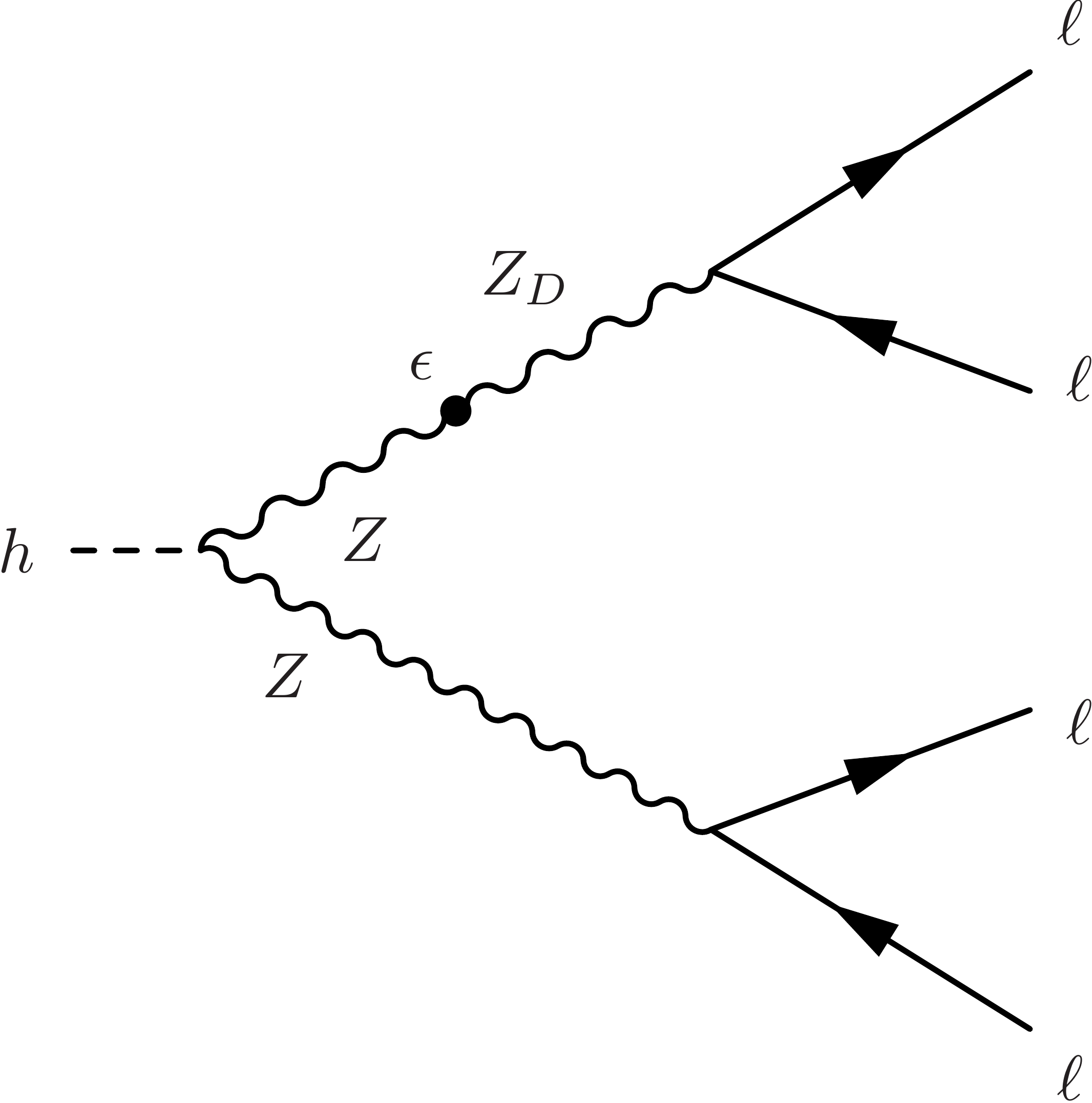}
    \includegraphics[width=0.33\textwidth]{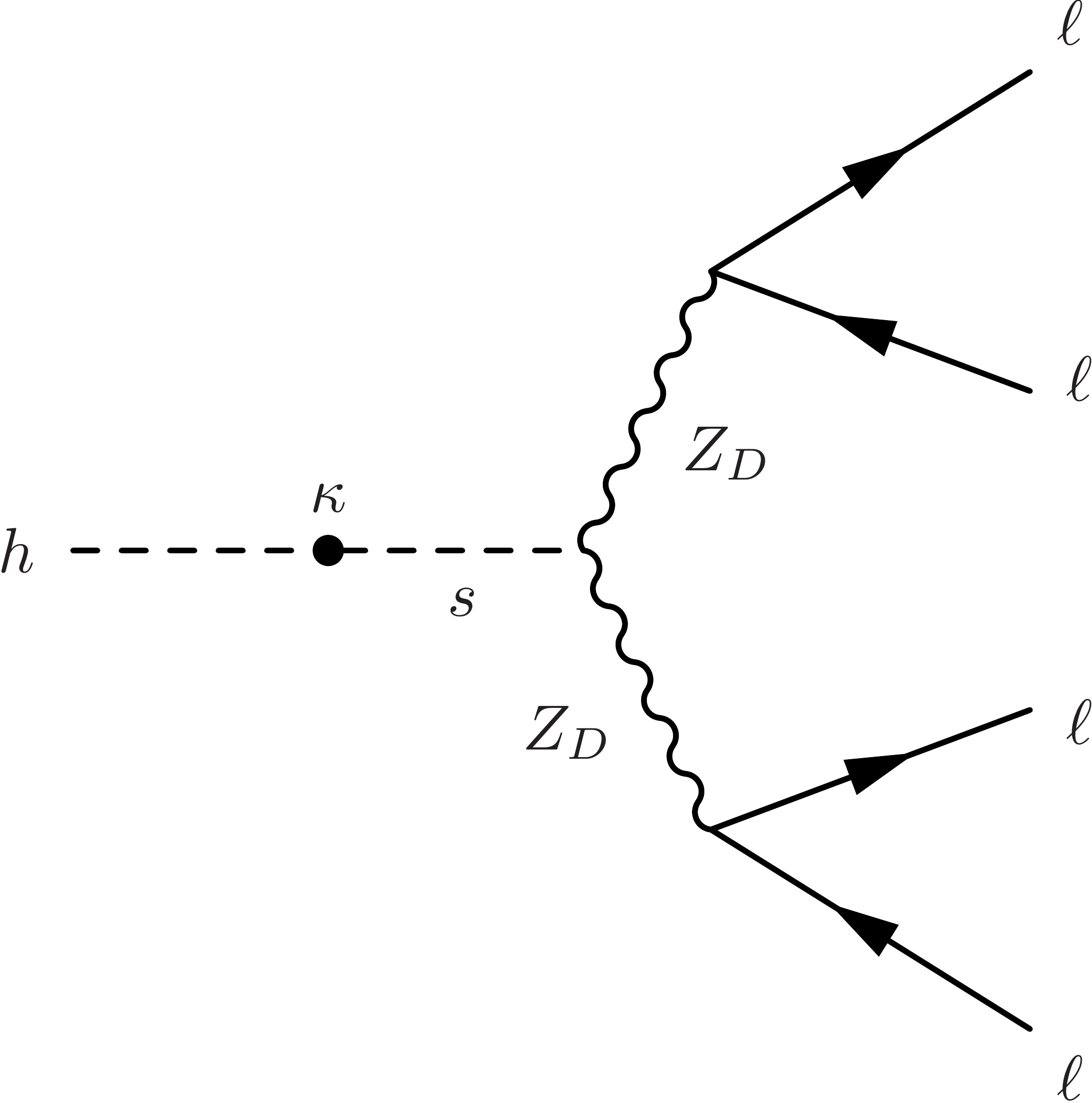}
    \caption{Feynman diagrams for Higgs boson decay via the kinetic-mixing (left) or Higgs-mixing mechanism (right)~\cite{Curtin:2014cca}. 
    The symbol $\Ph$ represents the Higgs boson, and $s$ represents the dark Higgs boson. The symbol \esp represents the kinetic-mixing 
    parameter while $\kappa$ represents the Higgs-mixing parameter.
    \label{fig:zdfeyn}}
\end{figure*}

The first model considered, hereby referred to as the ``hidden Abelian Higgs model'' (HAHM), concerns theories with a hidden ``dark'' sector~\cite{Curtin:2014cca,Curtin:2013fra,Davoudiasl:2013aya,Davoudiasl:2012ag,Gopalakrishna:2008dv}, with the X particle identified as the dark photon (\zd), which mediates a dark $U(1)_{D}$ gauge symmetry, which is spontaneously broken by 
a dark Higgs mechanism. Interactions of the dark sector with SM particles can occur through a hypercharge portal via the kinetic-mixing parameter \esp, or through a Higgs portal via the Higgs-mixing parameter $\kappa$, as shown in Fig.~\ref{fig:zdfeyn}. Details of this theory and subsequent phenomenological implications can be found in Ref.~\cite{Curtin:2014cca}. Several searches for \zd were previously performed by collider experiments, for example ATLAS~\cite{Aad:2015sva,Aaboud:2018fvk} and LHCb~\cite{Aaij:2017rft}. Other experiments, such as beam dump experiments, fixed target experiments, helioscopes, and cold dark matter direct detection experiments, provide complementary sensitivities to \zd. A summary of the experimental coverage of the HAHM model can be found in Refs.~\cite{Essig:2013lka,Beacham:2019nyx}.

{\tolerance=800
The second model involves axion-like particles (ALPs), with \PX being a pseudoscalar gauge singlet \Pa. 
Axions were originally proposed to address the strong CP problem~\cite{PhysRevLett.38.1440}.
Recently, ALPs were proposed to explain the observed anomaly in the magnetic moment of the muon~\cite{PhysRevLett.119.031802}. Theoretical 
overviews of the ALP models can be found in Refs.~\cite{Georgi:1986df,Bauer:2017ris}. The models are formulated as an effective field theory 
of ALPs coupled to various SM particles. In particular, the theory allows the coupling between the Higgs boson, Z boson, and the ALP field, or the Higgs 
boson and the ALP field. These couplings are represented by the Wilson coefficients \cZhLambda and \cahLambda, respectively, where $\Lambda$ is the decoupling energy scale in the 
effective field theory, or the mass scale of new physics. The former (latter) coefficient gives rise to the exotic decay of $\PH \to \PZ \Pa$ ($\Pa \Pa$). Various experimental searches for $\PH \to \Pa \Pa$ have been performed~\cite{Chatrchyan:2012cg, Aad:2015bua, Khachatryan:2015nba, Khachatryan:2017mnf,ATLAS:2021hbr,ATLAS:2021ldb}. 
Recently a direct search for $\PH \to \PZ \Pa$ has been performed targeting a signature with a light and hadronically decaying resonance $\Pa$ with ${\mass{\Pa} < 4\GeV}$~\cite{PhysRevLett.125.221802}. 
The present search provides complementary coverage of the phase space of the ALP model with mass greater than 4\GeV.
\par}

This paper is organized as follows. Section~\ref{sec:detector} describes the CMS detector and event reconstruction algorithms. 
Section~\ref{sec:dataset} outlines the collision data used and various software packages used to generate the samples of simulated
events. Section~\ref{sec:evtselection} summarizes the selection criteria and the categorization of signal events, and Section~\ref{sec:bkg} describes
the reducible background estimation method. Section~\ref{sec:sys} describes the various sources of systematic uncertainties in the
search. Finally, results and interpretations are detailed in Section~\ref{sec:result}, and a summary is given in Section~\ref{sec:summary}.
Tabulated results are provided in HEPDATA~\cite{hepdata}.

\section{The CMS detector and event reconstruction}
\label{sec:detector}

The central feature of the CMS apparatus is a superconducting solenoid of 6\unit{m} internal diameter, 
providing a magnetic field of 3.8\unit{T}. Within the solenoid volume are a silicon pixel and strip 
tracker, a lead tungstate crystal electromagnetic calorimeter (ECAL), and a brass and scintillator 
hadron calorimeter (HCAL), each composed of a barrel and two endcap sections. Forward calorimeters 
extend the pseudorapidity ($\eta$) coverage provided by the barrel and endcap detectors. Muons are detected 
in gas-ionization chambers embedded in the steel flux-return yoke outside the solenoid. A more 
detailed description of the CMS detector, together with a definition of the coordinate system used 
and the relevant kinematic variables, can be found in Ref.~\cite{Chatrchyan:2008zzk}.

Events of interest are selected using a two-tiered trigger system~\cite{Khachatryan:2016bia}. 
The first level, composed of custom hardware processors, uses information from the calorimeters 
and muon detectors to select events at a rate of around 100\unit{kHz} within a fixed time interval of about 4\mus. 
The second level, known as the high-level trigger, consists of a farm of processors running a version of 
the full event reconstruction software optimized for fast processing, and reduces the event rate to around 1\unit{kHz} 
before data storage.

{\tolerance=800
The candidate vertex with the largest value of summed physics-object $\pt^2$ (where \pt is the transverse momentum) is taken to be the primary $\Pp\Pp$ 
interaction vertex. The physics objects are the jets, clustered using the jet finding algorithm~\cite{Cacciari:2008gp,Cacciari:2011ma} 
with the tracks assigned to candidate vertices as inputs, and the associated missing transverse momentum, taken as the 
negative vector sum of the \pt of those jets.
\par}

The particle-flow (PF) algorithm~\cite{CMS-PRF-14-001} aims to reconstruct and identify each individual particle in an event (PF candidate), with an 
optimized combination of information from the various elements of the CMS detector. The energy of photons is obtained from the ECAL measurement. 
The energy of electrons is determined from a combination of the electron momentum at the primary interaction vertex as determined by the tracker, 
the energy of the corresponding ECAL cluster, and the energy sum of all bremsstrahlung photons spatially compatible with originating from the 
electron track. The energy of muons is obtained from the curvature of the corresponding track. The energy of charged hadrons is determined from 
a combination of their momentum measured in the tracker and the matching ECAL and HCAL energy deposits, corrected for the response function of 
the calorimeters to hadronic showers. Finally, the energy of neutral hadrons is obtained from the corresponding corrected ECAL and HCAL energies.

The missing transverse momentum vector \ptvecmiss is computed as the negative vector sum of the transverse momenta of all the PF 
candidates in an event, and its magnitude is denoted as \ptmiss~\cite{Sirunyan:2019kia}. The \ptvecmiss is modified to account for corrections 
to the energy scale of the reconstructed jets in the event.

Muons in the four lepton final state are measured in the range ${\abs{\eta} < 2.4}$, with detection planes made using three technologies: drift tubes, 
cathode strip chambers, and resistive plate chambers. The single-muon trigger efficiency exceeds 90\% over the full $\eta$ range, and the 
efficiency to reconstruct and identify muons is greater than 96\%. Matching muons to tracks measured in the silicon tracker results in a 
relative \pt resolution, for muons with \pt up to 100\GeV, of 1\% in the barrel and 3\% in the endcaps~\cite{Sirunyan:2018fpa}.

{\tolerance=800
Electrons in the four lepton final state with ${\pt > 7\GeV}$ and ${\abs{\eta} < 2.5}$ are identified by a multivariate discriminant, which 
is constructed by observables related to the bremsstrahlung along the electron trajectory, ECAL energy measurements, electromagnetic showers, missing pixel detector hits, 
and the photon conversion vertex fit probability~\cite{CMS:2020uim}.
The electron momentum is estimated by combining the energy measurement in the ECAL with the momentum measurement in the tracker. The momentum resolution for 
electrons with ${\pt \approx 45\GeV}$ from ${\PZ \to \Pe \Pe}$ decays ranges from 1.7 to 4.5\%. It is generally better in the barrel region than in the 
endcaps, and also depends on the bremsstrahlung energy emitted by the electron as it traverses the material in front of the ECAL.
The dielectron mass resolution for ${\PZ \to \Pe \Pe}$ decays when both electrons are in the ECAL barrel (endcap) is 1.9\% (2.9\%). 
\par}

This analysis focuses on promptly produced signal processes. To reduce the contributions from leptons arising from hadron decays within jets, a requirement is imposed on each lepton candidate using a variable defined as:
\begin{linenomath}
\begin{equation}
 I^{\ell} = \frac{\sum \pt^{\text{charged}}+\max\Bigl[0,\sum \pt^{\text{neutral}}+\sum \pt^{\PGg}-\pt^{\text{PU}}\Bigr]}{\pt^{\ell}}
\end{equation}
\end{linenomath}
where the sums are over the PF candidates within a cone of radius ${R = \sqrt{\smash[b]{\Delta\eta^2+\Delta\phi^2}} <  0.3}$ (where $\phi$ is the azimuthal angle in radians), $\pt^i$ represents transverse momenta from each particle $i$, where $i$ represents either charged hadrons, neutral hadrons, photons, or particles originating from overlapping proton-proton interactions (pileup)~\cite{Sirunyan:2017exp}. For muons, the isolation is required to be ${I^{\PGm} < 0.35}$.
For electrons, this variable is included in the multivariate discriminant for datasets in 2017 and 2018, while for the dataset in 2016, an isolation requirement ${I^{\Pe} < 0.35}$ 
is imposed on each electron candidate. In addition, the three-dimensional impact parameter of electrons and muons is required to be consistent with the primary collision vertex. 
The requirement implies a negligible acceptance to signal models with long-lived \PX.

An algorithm is utilized to correct for effects arising from final-state radiation (FSR) from leptons. PF-reconstructed photons are considered as FSR candidates if they satisfy the requirement ${\pt^{\PGg} > 2\GeV}$ and ${I^{\PGg} < 1.8}$, where $I^{\PGg}$ is calculated similarly to the lepton isolation variable. Then each FSR candidate is assigned to the closest lepton in the event. The candidates are further required to have ${\Delta R(\PGg,\ell)/(\pt^{\PGg})^2 < 0.012\GeV^{-2}}$ and ${\Delta R(\PGg,\ell) < 0.5}$. These candidates are excluded from the calculation of the lepton isolation variables.

Lepton reconstruction and selection efficiencies are measured in data by a ``tag-and-probe'' technique with an inclusive sample of \PZ boson events~\cite{Khachatryan:2010xn}. The difference between the efficiencies in data and simulation are observed to be around 1--4\%, depending on \pt and $\eta$ of the lepton considered. The differences are used to correct lepton efficiencies in simulation.

\section{Data and simulated samples}
\label{sec:dataset}

Leading order (LO) signal samples for the physics processes $\Pp\Pp \to \PH \to \PZ \zd (\zd \zd) \to 4\ell$, where $\ell = (\Pe,\PGm)$, are generated using the \MGvATNLO 2.2.2 (2.4.2)~\cite{MADGRAPH5,Alwall:2007fs,Frederix:2012ps}
generator for 2016 (2017 and 2018), with HAHM~\cite{Curtin:2014cca} at leading order.
Cross sections for each \zd signal are calculated by multiplying the next-to-next-to-next-to-leading order (NNNLO) Higgs production 
cross section~\cite{Anastasiou:2016cez} by the branching fraction of $\PH \to \PZ \zd$ and $\PH \to \zd \zd$,
respectively~\cite{Curtin:2014cca}. Final states with $\tau$ leptons are neglected as their contribution to the signal region yield 
is below $1\%$. Signal contributions from vector-boson fusion and associated production with a top quark pair or a vector boson are also 
omitted.

{\tolerance=800
The SM Higgs boson simulation samples, which include gluon fusion, vector boson fusion, and associated production with a top quark pair or a vector boson, and the simulated
$\PZ\PZ$ background from quark-antiquark annihilation are generated at next-to-leading order (NLO) in 
perturbative quantum chromodynamics with \POWHEG~v2~\cite{Bagnaschi:2011tu,Nason:2004rx,Frixione:2007vw,Alioli:2010xd}. 
The cross section for the dominant production mode, gluon fusion, is taken at NNNLO~\cite{Anastasiou:2016cez}.
\par} 

Decays of the Higgs boson to four leptons are simulated with \JHUGen 7.0.2~\cite{PhysRevD.81.075022,PhysRevD.86.095031}.
The non-resonant process of $\Pg\Pg\to \PZ\PZ$ process is simulated at LO with \MCFM 7.0.1~\cite{Campbell:2019dru}. 
NLO correction factors~\cite{Grazzini:2017mhc} are applied to the $\Pg\Pg\to \PZ\PZ$ process.

{\tolerance=800
Minor backgrounds from \ttV and triboson production processes are also simulated at LO and NLO, respectively, with the \MGvATNLO 2.2.2 (2.4.2)~\cite{MADGRAPH5,Alwall:2007fs,Frederix:2012ps}
generator for 2016 (2017 and 2018). 
\par}

{\tolerance=1200
The set of parton distribution functions (PDFs) used was NNPDF3.0~\cite{Ball:2014uwa} (NNPDF3.1~\cite{Ball:2017nwa}) for the 2016 (2017 and 2018) simulation.
Parton showering and hadronization are simulated using the \PYTHIA~8.230 generator~\cite{Sjostrand:2014zea}
with the CUETP8M1 (CP5) underlying event tune for the 2016 (2017 and 2018) simulation~\cite{Khachatryan:2015pea,Sirunyan:2019dfx}.
The response of the CMS detector is modeled using the \GEANTfour program~\cite{AGOSTINELLI2003250,1610988}.
Simulated events are reweighted according to a specified instantaneous luminosity and an average number of pileup events.
par}

\section{Event selection}
\label{sec:evtselection}

In the trigger system, events are required to have more than two leptons.
The overall trigger efficiency is measured in data using a sample of $4\ell$ events from single-lepton triggers and 
agreements are observed with simulation within 5\%, and is found to be larger than 99\%.

A set of requirements is applied to maximize the sensitivity of the search for a potential signal in the
\ZX and \XX event topologies. In both cases, at least four well-identified and isolated leptons from the primary
vertex are required, possibly accompanied by an FSR photon. Each muon (electron) is required to have $\pt > 5\GeV$ (7\GeV). All four leptons must be separated from each
other by $\Delta R(\ell_i,\ell_j) > 0.02$. The leading (subleading) lepton \pt is required to satisfy $\pt > 20\GeV$ (10\GeV).
The four-lepton invariant mass $m_{4\ell}$ is required to be within $118 < m_{4\ell} < 130\GeV$. 
To further suppress background contributions from hadron decays in jet fragmentation or from the decay of low-mass 
resonances, all opposite-charge leptons pairs, regardless of lepton flavor, are required to satisfy
$m_{\ell^+ \ell^-} > 4\GeV$. 

For each event in the \ZX and \XX searches, dilepton pair candidates are formed by considering all OSSF leptons. 
The dilepton invariant mass $m_{\ell^{+} \ell^{-}}$ for each candidate is required to be within 
$4 < m_{\ell^{+}\ell^{-}} < 120\GeV$, however the mass window around the $\Upsilon \PQb\PAQb$ bound states ($8.0 < m_{\Upsilon} < 11.5\GeV$) is also excluded.

{\tolerance=800
Two dilepton candidates are then paired to form a \ZX or \XX event candidate. 
For the \ZX search, \ZOne is the OSSF dilepton pair with an
invariant mass closest to the \PZ boson mass~\cite{10.1093/ptep/ptaa104} (representing \PZ in \ZX), and 
\ZTwo is the other pair (\PX). For the \XX search, \ZOne is the OSSF dilepton pair with
the larger invariant mass, and \ZTwo is the lower-mass pair. For the \ZX search, $\mass{\ZOne}$ is required
to be larger than 40\GeV. For the \XX search, $\mass{\ZOne}$ and $\mass{\ZTwo}$ must lie between 4 and  62.5\GeV. 
For events with more than four selected leptons, the combination of four leptons with $\mass{\ZOne}$ closest to the Z boson 
is used for the \ZX candidate, while the combination with the least value of $(\mass{\ZOne}-\mass{\ZTwo})/(\mass{\ZOne}+\mass{\ZTwo})$ is used to select \XX candidates with similar invariant masses. 
\par}

Four final-state lepton categories can be defined as \fourMu, \twoMuTwoEl, \fourEl, \twoElTwoMu, 
where the order of lepton flavors corresponds to Z1 and Z2 flavors.
For the \fourMu and \fourEl final states, one alternative pairing of the 
four leptons is possible, labelled by $\PZ_\text{a}$ and
$\PZ_\text{b}$. 
For the \ZX search, events with $m_{\PZ_\text{b}} < 12\GeV$ and $m_{\PZ_\text{a}}$
closer to the \PZ boson mass than \ZOne are discarded to suppress background 
contributions from on-shell Z and low-mass dilepton resonances.
For the \XX search, the \XX candidate with the smallest value of 
$(\mass{\ZOne}-\mass{\ZTwo})/(\mass{\ZOne}+\mass{\ZTwo})$ is chosen.

\section{Background estimation}
\label{sec:bkg}
\subsection{Irreducible background estimation}
Irreducible backgrounds for this search come from processes including a SM Higgs boson, as well as nonresonant production of $\PZ\PZ$ via quark-antiquark annihilation or gluon fusion, 
and rare backgrounds such as $\PQt\PAQt+\PZ$ and triboson production. These backgrounds are estimated from simulation. Details of the simulation used for each of the 
backgrounds are described in Section~\ref{sec:dataset}.

\subsection{Reducible background estimation}
\label{sec:zx}

The reducible backgrounds in the $4\ell$ final state can arise from the leptonic decays of heavy-flavor hadrons, 
in-flight decays of light mesons within jets, charged hadrons misidentified as electrons when in proximity 
of a $\PGpz$, and photon conversions. These backgrounds primarily arise from the \zj process. Additional physics processes 
with kinematics similar to the signal include \ttNew, \zg, and \wz. 

Two dedicated control regions are used to estimate the contribution from these backgrounds. 
The first (second) control region consists of events with two (three) leptons passing the lepton identification and isolation requirements 
and two (one) leptons failing the requirements, and is denoted as the 2P2F (3P1F) region. 
Backgrounds with only two prompt leptons, such as \zj and \ttNew, are estimated by the 2P2F region, 
while backgrounds with three prompt leptons, such as \wz and \zg with the photon converting to an electron pair, 
are estimated by the 3P1F region. Other than the lepton requirements, the 3P1F and 2P2F regions follow the same event selection and alternative pairing 
algorithms as in the signal region to closely mimic its kinematics.

The lepton misidentification rates $f_{\PGm}$ and $f_{\Pe}$ are measured as a function of lepton \pt and $\eta$ with a sample which 
includes a \PZ candidate, formed by a pair of leptons passing the selection requirement of the analysis, and an additional lepton passing 
a relaxed requirement. These rates are measured separately in the data samples from 2016, 2017, and 2018. In addition, the mass of the \PZ candidate is required to satisfy the condition $\abs{\mass{\ZOne}-\mass{\PZ}} < 7\GeV$ to reduce contributions 
from \wz and \ttNew processes, and \ptmiss is required to be less than 25\GeV.

To estimate the background contribution in the signal region, events in the 3P1F and 
2P2F control regions are reweighted by lepton misidentification probabilities. 
Each event $i$ in the 3P1F region is weighted by a factor $f^{i}_{4}/(1-f^{i}_{4})$, 
where $f^{i}_{4}$ corresponds to the lepton misidentification rate of the failed lepton in the event.
Physics processes in the 2P2F control region can contribute to the 3P1F region 
and are estimated by reweighting 2P2F events with $f^{i}_{3}/(1-f^{i}_{3})+f^{i}_{4}/(1-f^{i}_{4})$, 
where $f^{i}_{3}$ and $f^{i}_{4}$ correspond to the lepton misidentification rates of 
the two failed leptons in the event. A minor contribution from $\PZ\PZ$ events to the 3P1F control region is estimated from simulation and subtracted.
The expected yield for the signal region can then be estimated as:
\begin{linenomath}
\ifthenelse{\boolean{cms@external}}
{ 
\begin{multline}
    N^{\text{reducible}}_{\mathrm{SR}} = \left( 1-\frac{N^{\PZ\PZ}_{3P1F}}{N_{3P1F}} \right) \\
    \times \sum_{i}^{N_{3P1F}} \frac{f^{i}_{4}}{1-f^{i}_{4}} -
    \sum_{i}^{N_{2P2F}} \frac{f^{i}_{3}}{1-f^{i}_{3}} \frac{f^{i}_{4}}{1-f^{i}_{4}}
\end{multline}
} 
{ 
\begin{equation}
    N^{\text{reducible}}_{\mathrm{SR}} = \left( 1-\frac{N^{\PZ\PZ}_{3P1F}}{N_{3P1F}} \right) \sum_{i}^{N_{3P1F}} \frac{f^{i}_{4}}{1-f^{i}_{4}} - \sum_{i}^{N_{2P2F}} \frac{f^{i}_{3}}{1-f^{i}_{3}} \frac{f^{i}_{4}}{1-f^{i}_{4}}
\end{equation}
} 
\end{linenomath}
where each sum is over all 3P1F and 2P2F events, respectively.

Furthermore, dedicated validation regions, which include adjacent $m_{4\ell}$ regions to the signal region ($70 < m_{4\ell} < 118~\GeV$, $130 < m_{4\ell} < 200~\GeV$), are defined to inspect the level of agreement between data and predictions.

\section{Systematic uncertainties}
\label{sec:sys}

{\tolerance=800
Experimental sources of the systematic uncertainties applicable to all final states include the integrated luminosity uncertainty and the lepton identification and reconstruction efficiency 
uncertainty. The integrated luminosities of the 2016, 2017, and 2018 data-taking periods are individually known with uncertainties in the 1.2--2.5\% 
range~\cite{CMS-PAS-LUM-17-001,CMS-PAS-LUM-17-004,CMS-PAS-LUM-18-002}, while the total Run~2 (2016--2018) integrated luminosity has an uncertainty of 1.6\%~\cite{CMS:2021xjt}, the improvement in precision 
reflecting the (uncorrelated) time evolution of some systematic effects. 
Lepton efficiency uncertainties are estimated in bins of lepton \pt and $\eta$ using the tag-and-probe 
method, as described in Section~\ref{sec:detector}. These uncertainties on each lepton candidate lead to variations from 2.5 to 16.1\% on event yields, dependent on final-state lepton categories.  
In addition, the systematic uncertainties in the lepton energy scale are determined by fitting the $\PZ\to\ell\ell$ mass distribution in bins of 
lepton \pt and $\eta$ with a Breit--Wigner parameterization convolved with a double-sided Crystal Ball function~\cite{Oreglia:1980cs}.  
Systematic uncertainties in the estimation of the reducible background are derived from the level of agreement between data and predictions in the validation regions in each lepton 
category (23--48\% depending on data taking period), arising from different background compositions between signal and control regions (30--38\% depending on lepton category), 
and from misidentification rate uncertainties (35--100\% depending on lepton category).
\par}

Theoretical uncertainties that affect both the signal and background estimation include uncertainties in the renormalization and factorization scales and the choice of the PDF set. 
The uncertainty from the renormalization and factorization scales is determined by varying these scales between 0.5 and 2 times their nominal value while keeping their ratio between 0.5 and 2. 
The uncertainty from the PDF set is determined by taking the root-mean-square of the variation when using different replicas 
of the default NNPDF set~\cite{Butterworth:2015oua}. An additional uncertainty of 10\% in the K factor used for the $\Pg\Pg\to4\ell$ prediction is included~\cite{Sirunyan:2017exp}. 
To estimate the effect of the interference between the signal and background processes, three types of samples are generated using the \MGvATNLO 2.4.2~\cite{MADGRAPH5,Alwall:2007fs,Frederix:2012ps}
generator: inclusive sample (${\PH \to \PZ \PZ^{*} \to 4\ell}$, ${\PH \to \PZ \PX / \PX \PX \to 4\ell}$), signal-only sample ${\PH \to \PZ \PX / \PX \PX \to 4\ell}$ and background-only sample ${\PH \to \PZ \PZ^{*} \to 4\ell}$. The inclusive sample contains background, signal, and interference contributions. The effect of the interference on the normalization of the signal is estimated by taking the difference of the inclusive sample cross section and the sum of the cross sections of the signal and background samples. This difference is at 1-2\% after the final event selection.
Theoretical values of branching fractions $\mathcal{B}(\zd \to \Pe\Pe\ \text{or}\ \PGm\PGm)$ are calculated in Ref.~\cite{Curtin:2014cca}.
The calculations are based on experimental measurements of the ratio of the hadronic cross section to the muon cross section in electron-positron collisions $R_{\PGm\PGm}/R_{\text{had}}$ up to $\mass{\zd} = 12\GeV$ and a next-to-leading
order theoretical calculation for $\mass{\zd} > 12\GeV$. To account for uncertainties in these theoretical estimates, a conservative
20\%\,(10\%) uncertainty is assigned to them for $\mass{\zd} < 12\GeV$ ($\mass{\zd} > 12\GeV$)~\cite{Curtin:2014cca}.
Differences in the kinematic properties between the HAHM and ALP model have been inspected. For the determination of model-independent exclusion limits, differences in acceptances are included as systematic uncertainties, ranging from 10\% ($\mass{\PX} \sim 4\GeV$) to 30\% ($\mass{\PX} \sim 35\GeV$ for \ZX, $\mass{\PX} \sim 60\GeV$ for \XX), while they are used to correct signal yields for the determination of ALP exclusion limits.

In the combination of the three data taking periods, the theoretical uncertainties and experimental ones related to 
leptons are correlated across all data taking periods, while all others from experimental sources are taken as uncorrelated. The sensitivity of this analysis is dominated by data statistical 
uncertainty rather than systematic uncertainties.

\section{Results and interpretation}
\label{sec:result}

\begin{figure}[!htbp]
\centering
\includegraphics[width=0.49\textwidth]{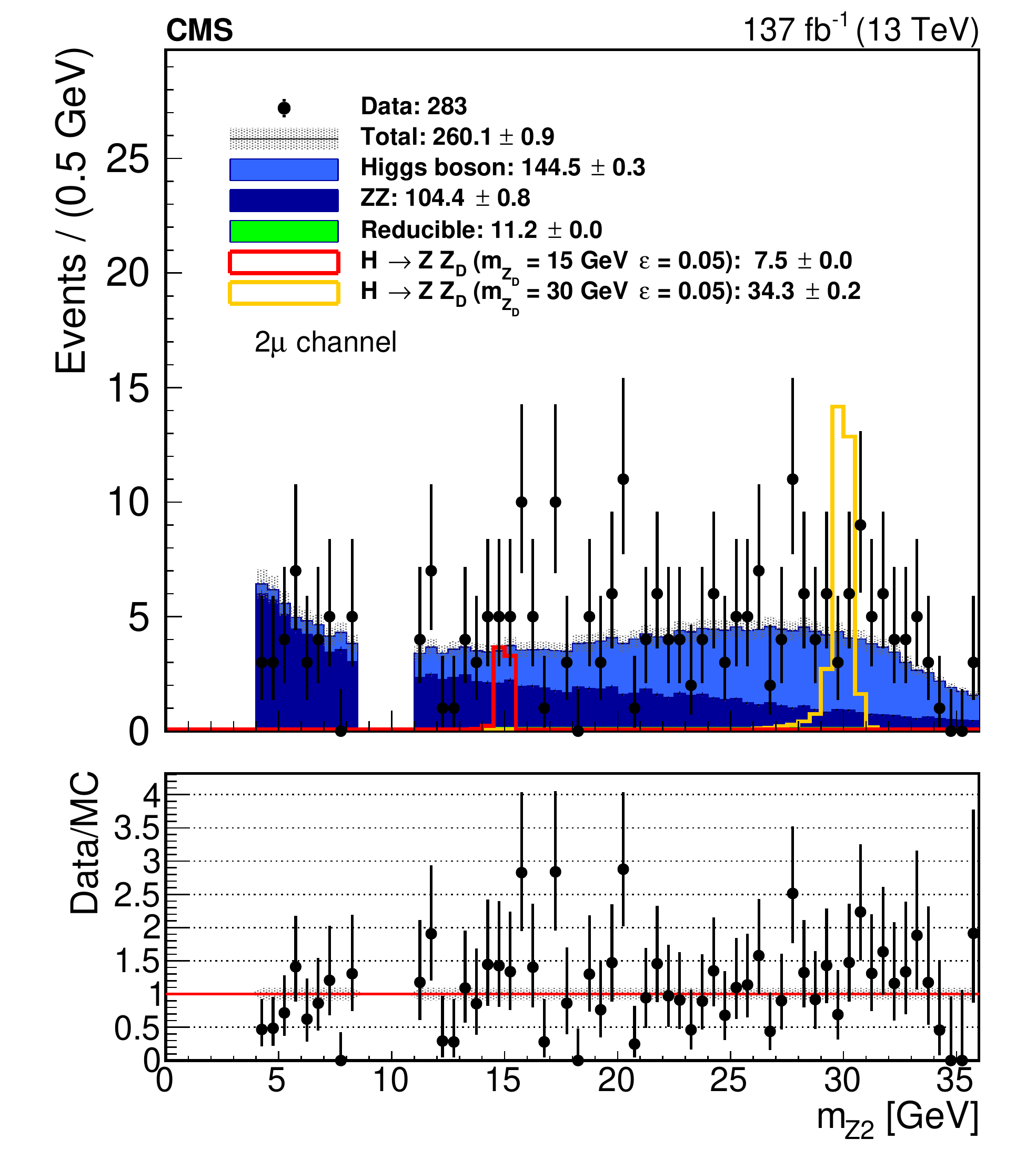}
\includegraphics[width=0.49\textwidth]{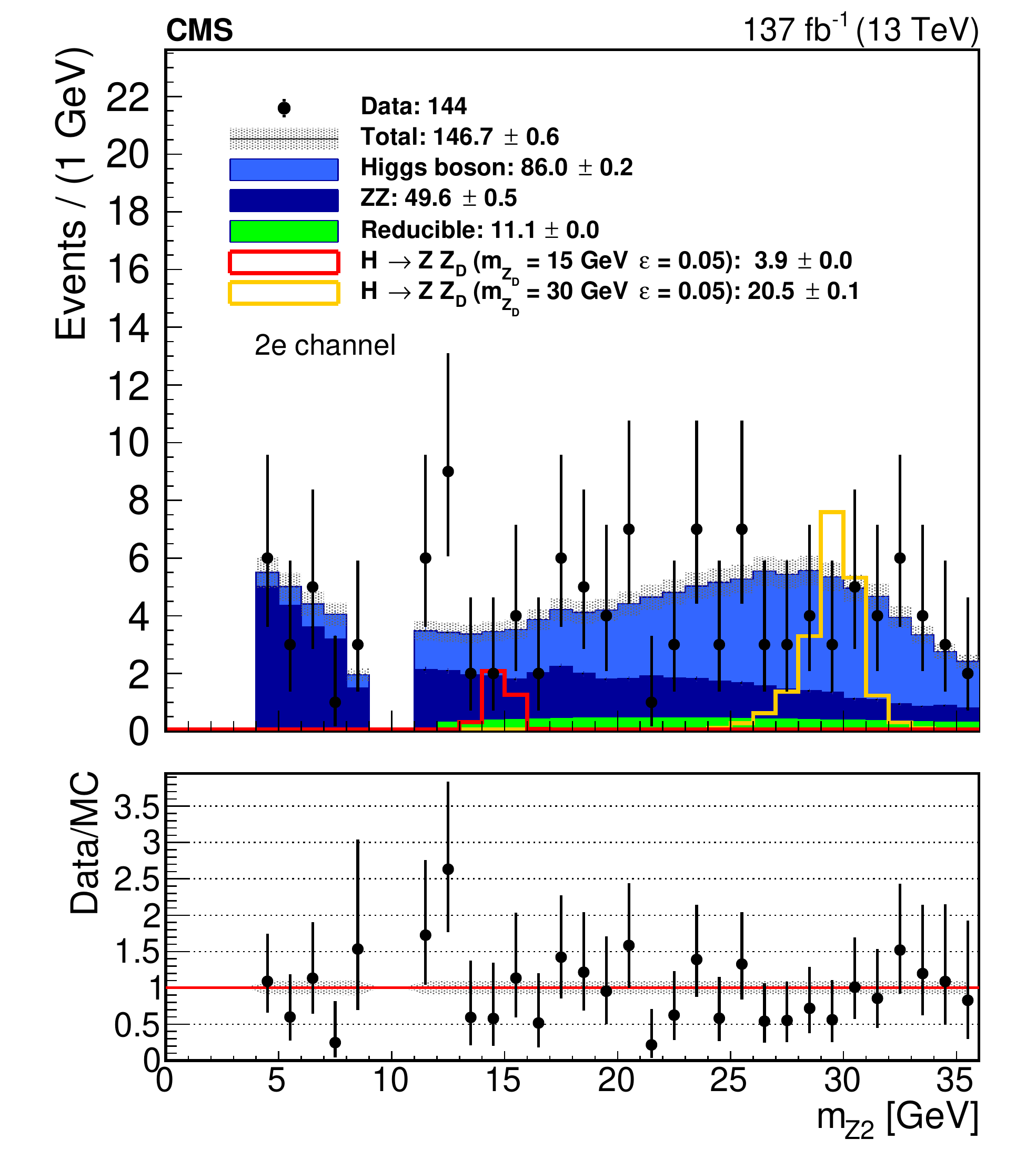}
    \caption{Event yields against $m_{\PZ_2}$ with the \ZX selection for the muon and electron channels. 
    Numbers in the legend show the total event yields with the \ZX selection corresponding to 
    data, and the expected yields for each background and signal processes, along with the corresponding 
    statistical uncertainty coming from the amount of simulated data. \label{fig:ZZd_mZ2}}
\end{figure}

\begin{figure*}[!htb]
\centering
\includegraphics[width=\cmsFigWidth]{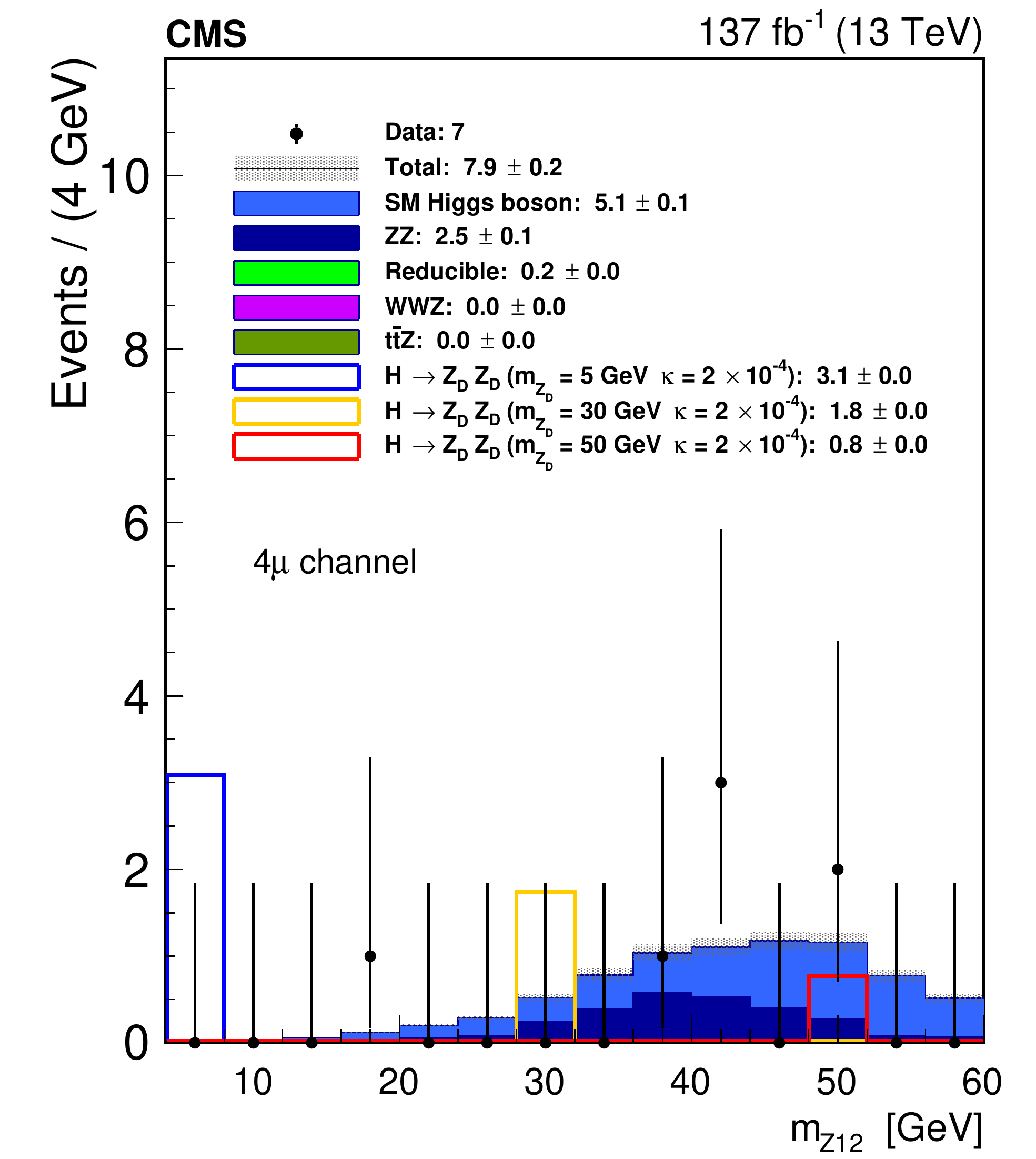}
\includegraphics[width=\cmsFigWidth]{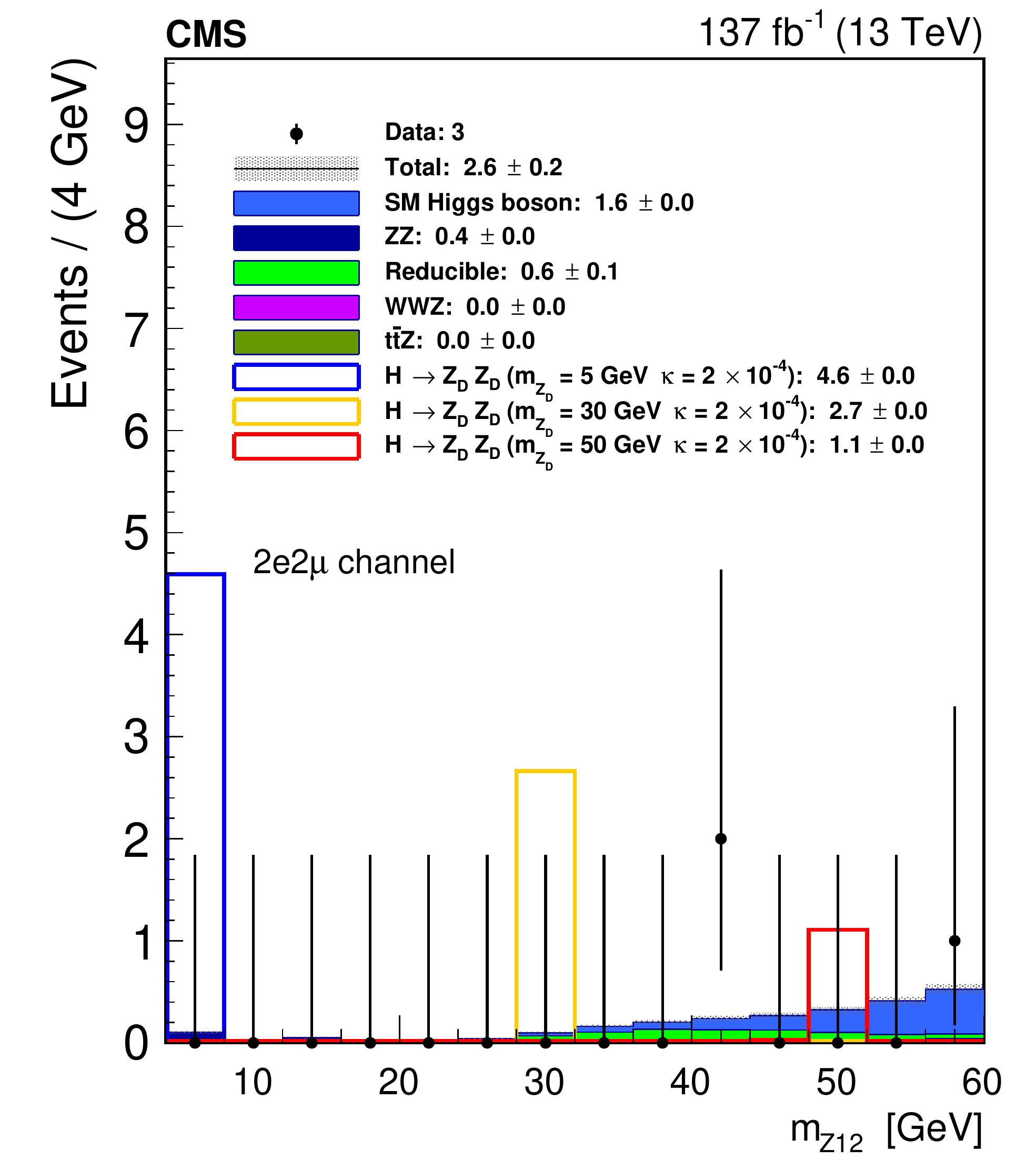}
\includegraphics[width=\cmsFigWidth]{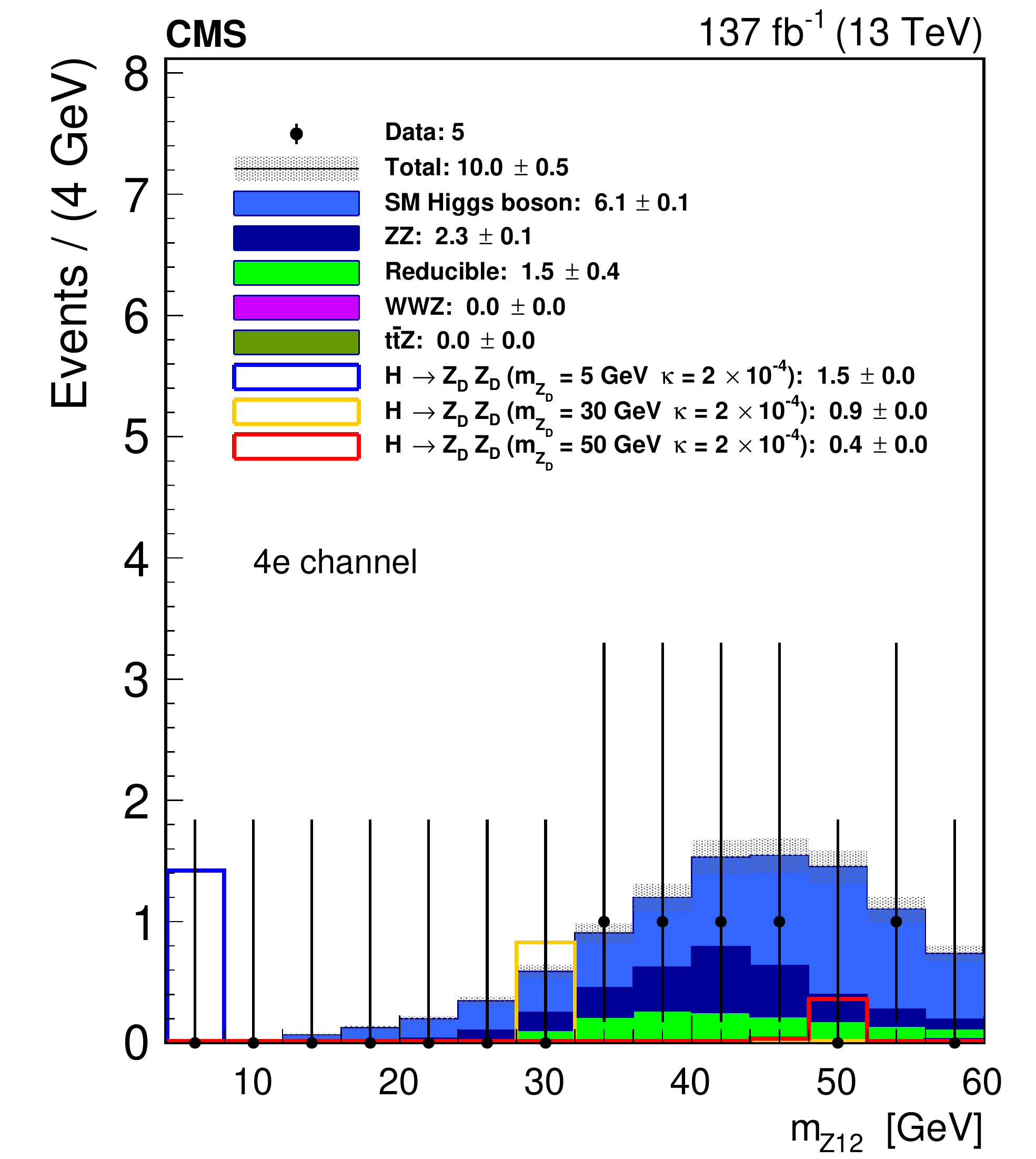}
    \caption{Event yields against $\mass{Z12} = (\mass{\ZOne}+\mass{\ZTwo})/2$ with the \XX selection for the 
    \fourMu, \twoElTwoMu and \fourEl final states. Numbers in the legend show the total event yields 
    with the \XX selection corresponding to data, and the expected yields for each background and signal processes,
    along with the corresponding statistical uncertainty coming from the amount of simulated data.\label{fig:ZdZd_mZ12}}
\end{figure*}

Dilepton mass distributions for the \ZX and \XX selections are shown in
Figs.~\ref{fig:ZZd_mZ2} and~\ref{fig:ZdZd_mZ12}, respectively.  
The dilepton mass variable for the \XX selection shown in Fig.~\ref{fig:ZdZd_mZ12}
is $\mass{Z12} = (\mass{\ZOne}+\mass{\ZTwo})/2$, which should peak at $\mass{\PX}$ in case of a signal $\PH \to \PX \PX$.
In all cases, the observed distributions agree well with standard model expectations within the assigned uncertainties. 

These results are further interpreted as upper limits on model-independent branching fractions and model parameters for  
the dark photon and ALP models. For interpretations of the results of the \ZX selection, 351 mass hypotheses are considered. Each mass  
hypothesis $m_{i}$ is defined with an incremental step of 0.5\%, as $m_{i} = 4.20 \times 1.005^{i}$,
where $i = 0,1,2,\ldots,424$, excluding the mass points around the $\Upsilon$ $\PQb\PAQb$ bound states between $8.0 < m_{\Upsilon} < 11.5\GeV$ ($i = 130,131,\ldots,201$).
The incremental step is chosen so as not to miss any potential signal contribution due to detector resolution in \mass{\ZTwo}. For each 
mass hypothesis, the counting experiments are performed on the \mass{\ZTwo} distribution, with the bin centered at each mass hypothesis.
Because of the finite mass resolution of \mass{\ZTwo}, the choice of the bin width needs to be defined such that most of the signal contribution is included 
in the bin. The bin width is defined as $0.04\,(0.10) \times m_{i}$ for the \fourMu and \twoElTwoMu (\fourEl and \twoMuTwoEl) categories. 
This width is chosen as two times the \mass{\ZTwo} resolution and includes $\approx$95\% of signal events.
The normalization of the Higgs background is allowed to float freely in the likelihood fit. For each mass hypothesis, 
events outside the mass window are included as a sideband to constrain the normalization parameter.
No significant deviation with respect to the SM prediction is observed.

For interpretations of the results of the \XX selection, 
462 mass hypotheses are considered instead. In contrast to the \ZX interpretations, the counting experiments are performed by constructing
a rectangular region, centered at each mass hypothesis, in the (\mass{\ZOne},\mass{\ZTwo}) plane. The rectangular regions are effectively 
triangular as \mass{\ZOne} is defined as the larger invariant mass. The bin widths are defined in a similar manner as 
$0.04 m_{i}$ ($0.10 m_{i}$) for \mass{\ZOne} or \mass{\ZTwo} formed by muon (electron) pairs.

The likelihood model for each mass hypothesis is formulated as
\begin{equation}
\mathcal{L}_{m}  
    = \mathcal{L}_{m,\mathrm{SR}}  \mathcal{L}_{m,\text{SB}}
\end{equation}
\begin{linenomath}
\ifthenelse{\boolean{cms@external}}
{ 
\begin{multline}
\mathcal{L}_{m,\mathrm{SR}} = \prod_{\ell} \Pois ( n_{m,\ell} | \mu_{\text{Higgs}}  n_{\text{Higgs},m,\ell}  \\
    +\sum_{b} n_{b,m,\ell}  \rho_{b,m,\ell} + \mu  n_{s,m,\ell}  \rho_{s,m,\ell})
\end{multline}
} 
{ 
\begin{equation}
\mathcal{L}_{m,\mathrm{SR}} = \prod_{\ell} \Pois ( n_{m,\ell} | \mu_{\text{Higgs}}  n_{\text{Higgs},m,\ell} + \sum_{b} n_{b,m,\ell}  \rho_{b,m,\ell} + \mu  n_{s,m,\ell}  \rho_{s,m,\ell}),
\end{equation}
} 
\end{linenomath}
\begin{equation}
\label{eq:sb}
\mathcal{L}_{m,\text{SB}}
    = \prod_{\ell} \Pois ( n_{\ell} | \mu_{\text{Higgs}}  n_{\text{Higgs},\ell} + \sum_{b} n_{b,\ell}  \rho_{b,\ell} ) 
\end{equation}
where the function $\Pois(n|x)$ is the Poisson probability to observe $n$ events, when the expectation is $x$. 
The symbol $m$ represents a particular mass hypothesis. The likelihood term $\mathcal{L}_{m,\mathrm{SR}}$ ($\mathcal{L}_{m,\text{SB}}$) 
corresponds to the event yields within (outside) the mass window.
The symbol $\mu$ is the signal strength parameter, $\mu_{\text{Higgs}}$ represents the free floating normalizing parameter on the SM Higgs boson process, 
$\ell$ represents each lepton category, 
$b$ represents each background process, $s$ represents a particular signal process and $n_{i,m,\ell}$ represents the yield in a mass window associated with the mass hypothesis $m$, from a source $i$ and the lepton category $\ell$.
In Equation~\ref{eq:sb}, the symbols $n_{\text{Higgs},\ell}$ and $n_{b,\ell}$ represent the yields of the SM Higgs boson and other backgrounds $b$ outside the mass window for the lepton category $\ell$.
Systematic uncertainties are included and profiled as nuisance parameters $\rho$~\cite{ATL-PHYS-PUB-2011-011}.

For each interpretation, $95\%$ exclusion limits are obtained with an asymptotic formulation of the modified frequentist \CLs 
criterion as described in Refs.~\cite{Junk:1999kv,Read:2002hq,ATL-PHYS-PUB-2011-011,Cowan:2010js} with the ZX selection and 
full \CLs approach for the XX selection.

\subsection{Model-independent limits}
\begin{figure}[htb!p]
\centering
\includegraphics[width=\cmsFigWidth]{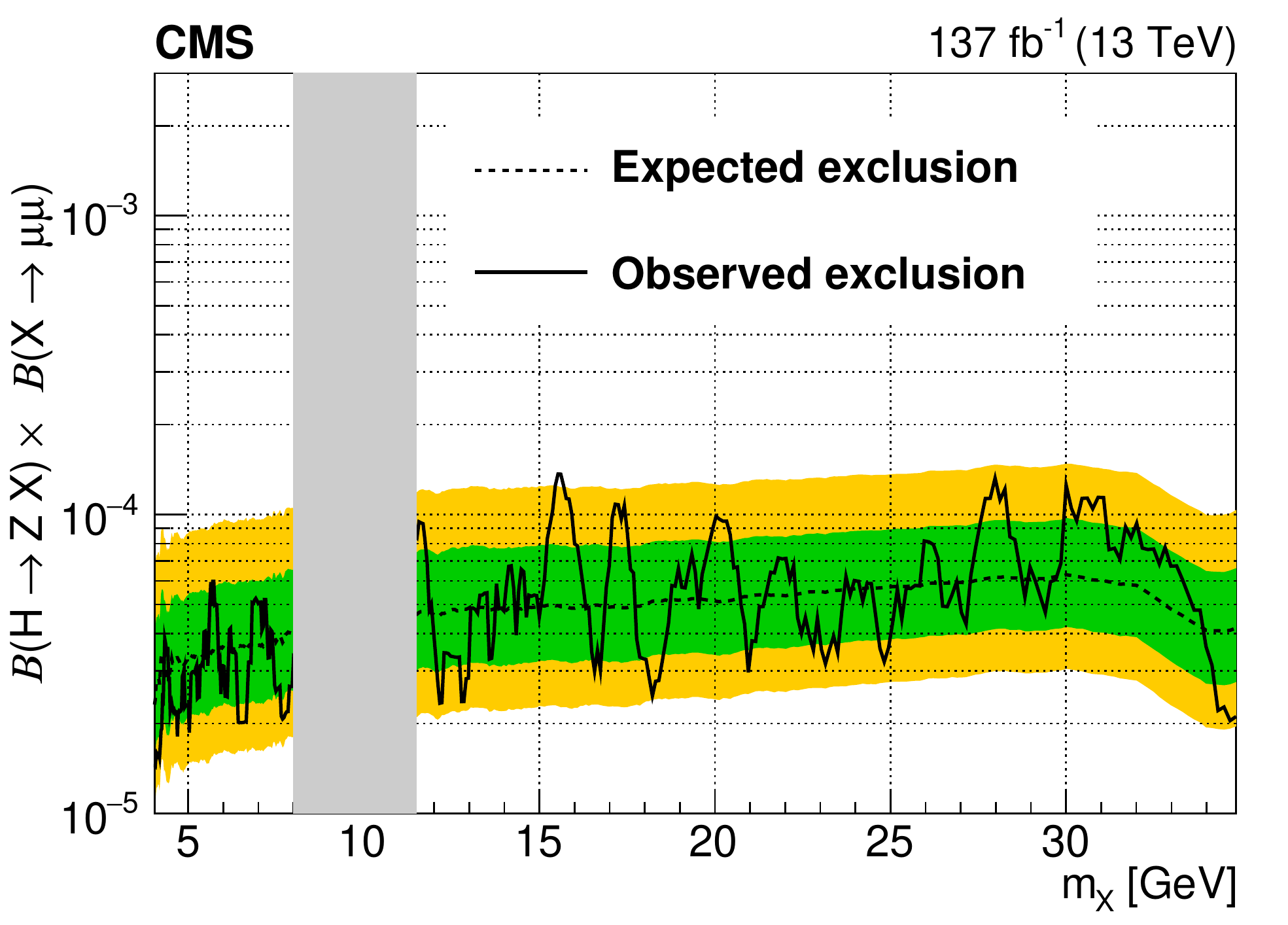}
\includegraphics[width=\cmsFigWidth]{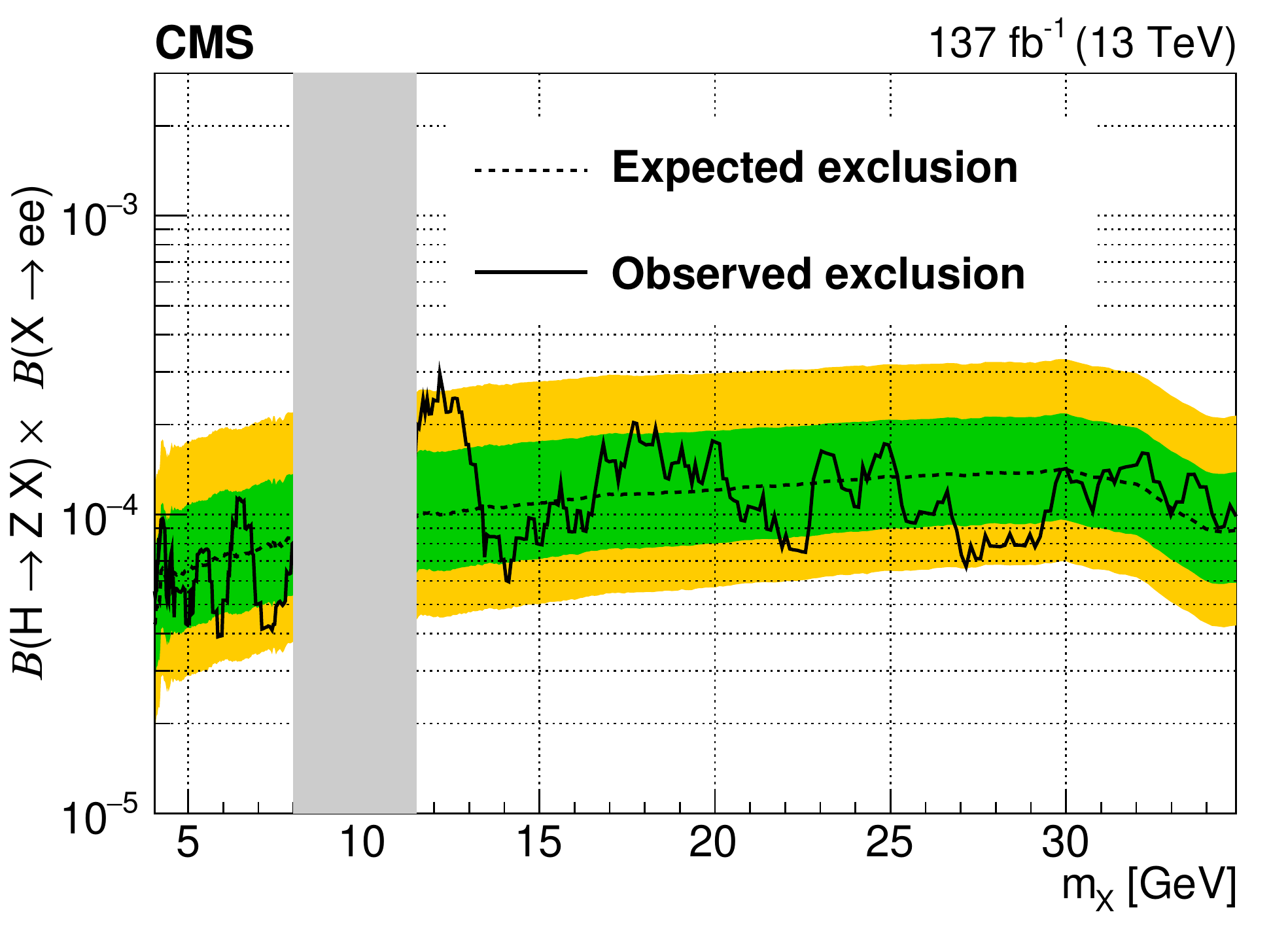}
\includegraphics[width=\cmsFigWidth]{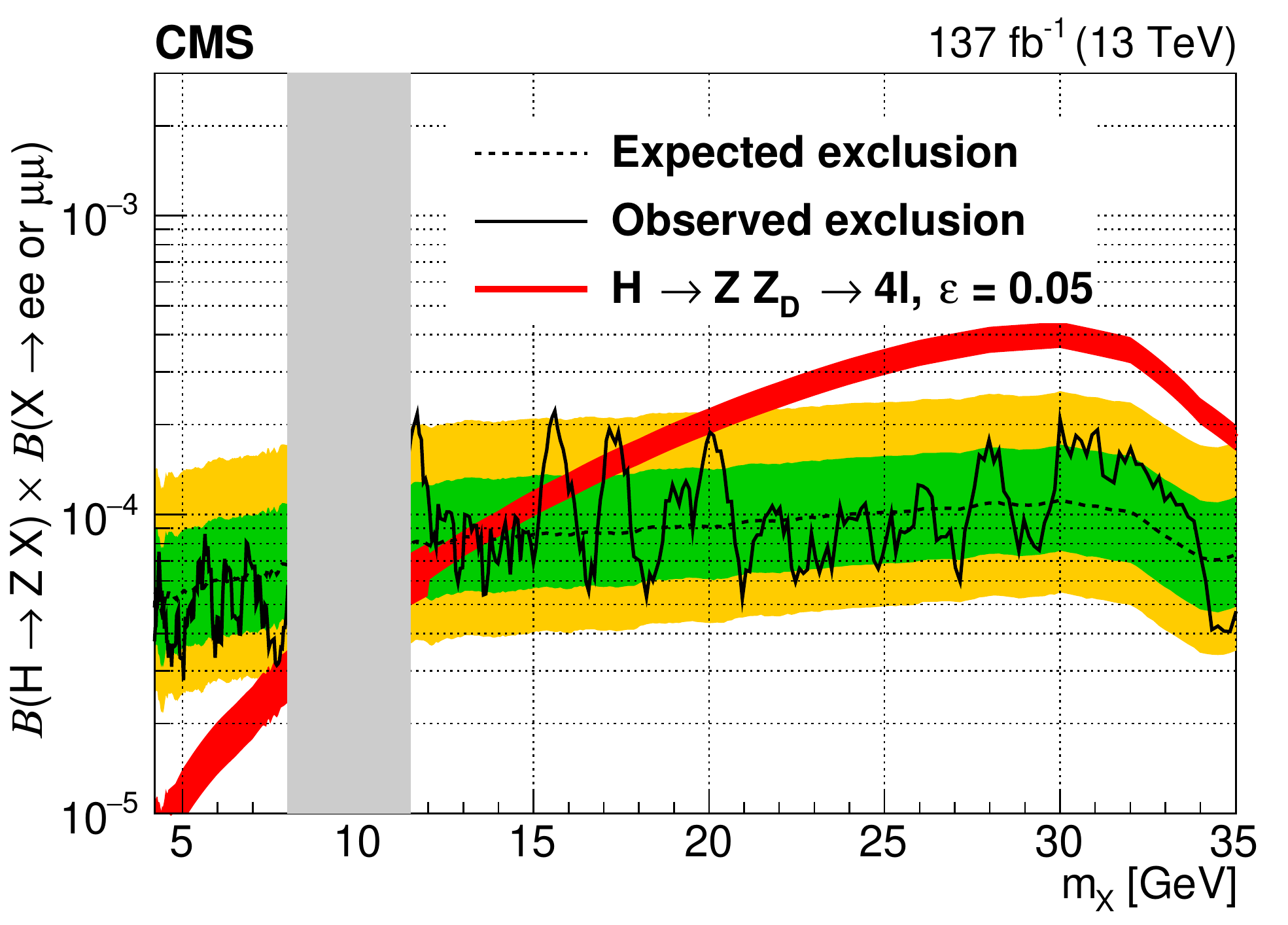}
\caption{
    Expected and observed 95\% CL limits on 
    $\mathcal{B}(\PH \to \PZ \PX)  \mathcal{B}(\PX \to \PGm\PGm)$ assuming \PX decays to dimuons only, 
    $\mathcal{B}(\PH \to \PZ \PX)  \mathcal{B}(\PX \to \Pe\Pe)$ assuming \PX decays to dielectrons only, 
    and $\mathcal{B}(\PH \to \PZ \PX)  \mathcal{B}(\PX \to \Pe\Pe\ \text{or}\ \PGm\PGm)$ assuming a flavor symmetric decay of \PX to dimuons and dielectrons.
    The dashed black curve is the expected upper limit, with one and two standard-deviation bands shown in green and yellow, respectively.
    The solid black curve is the observed upper limit. 
    The red curve represents the theoretical cross section for the signal process $\PH \to \PZ \PX \to 4\ell$. 
    The discontinuity at 12\GeV in the uncertainty is due to 
    the switch from experimental to theoretical uncertainty estimates of $\mathcal{B}(\zd \to \Pe\Pe\ \text{or}\ \PGm\PGm)$, as described in Ref.~\cite{Curtin:2014cca}. 
    The symbol $\esp$ is the kinetic-mixing parameter. 
    The grey band corresponds to the excluded region around the $\PQb\PAQb$ bound states of $\Upsilon$.
    \label{fig:limit_Br_ZX}
    }
\end{figure}

\begin{figure}[htb!p]
\centering
\includegraphics[width=\cmsFigWidth]{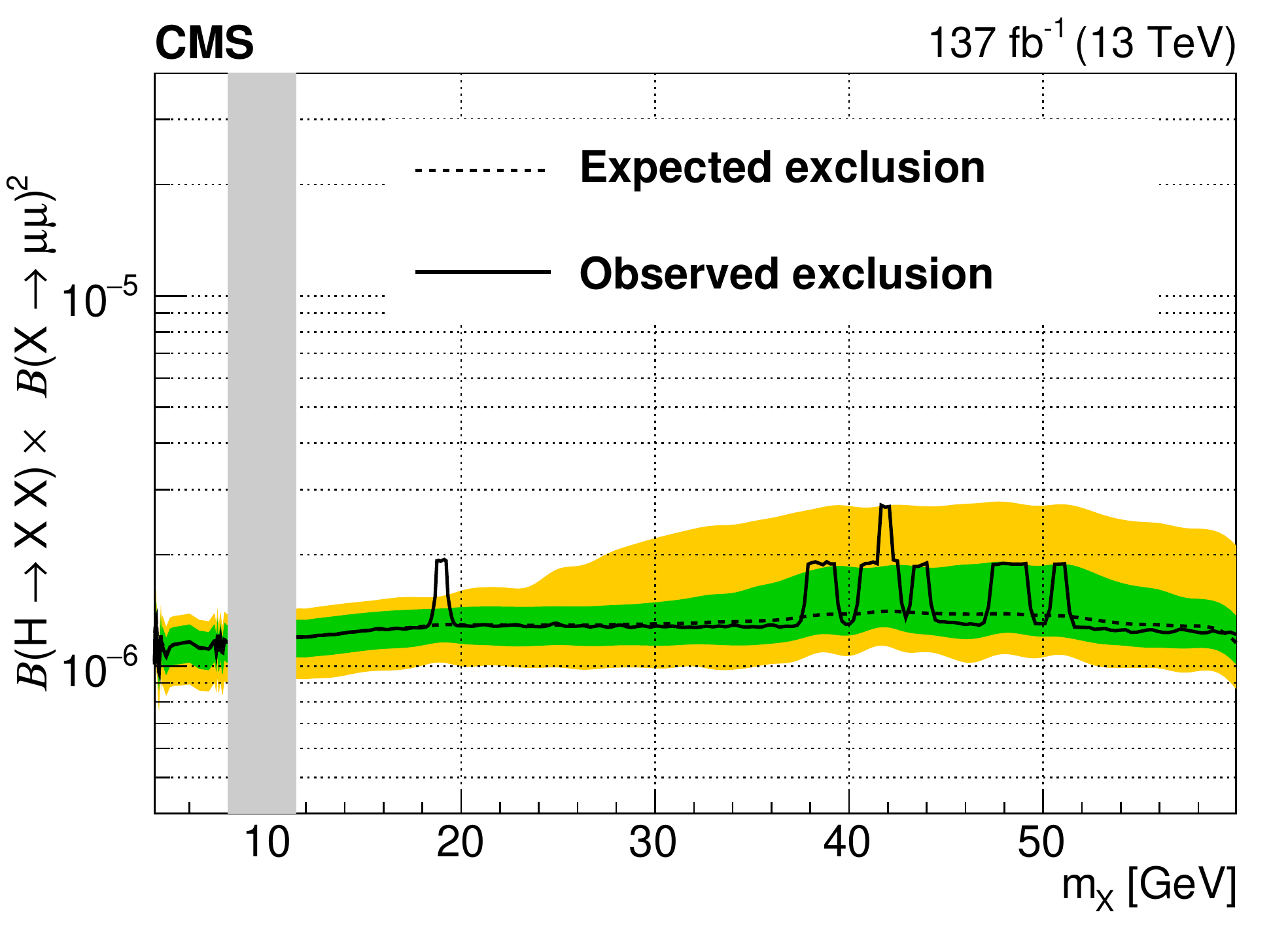}
\includegraphics[width=\cmsFigWidth]{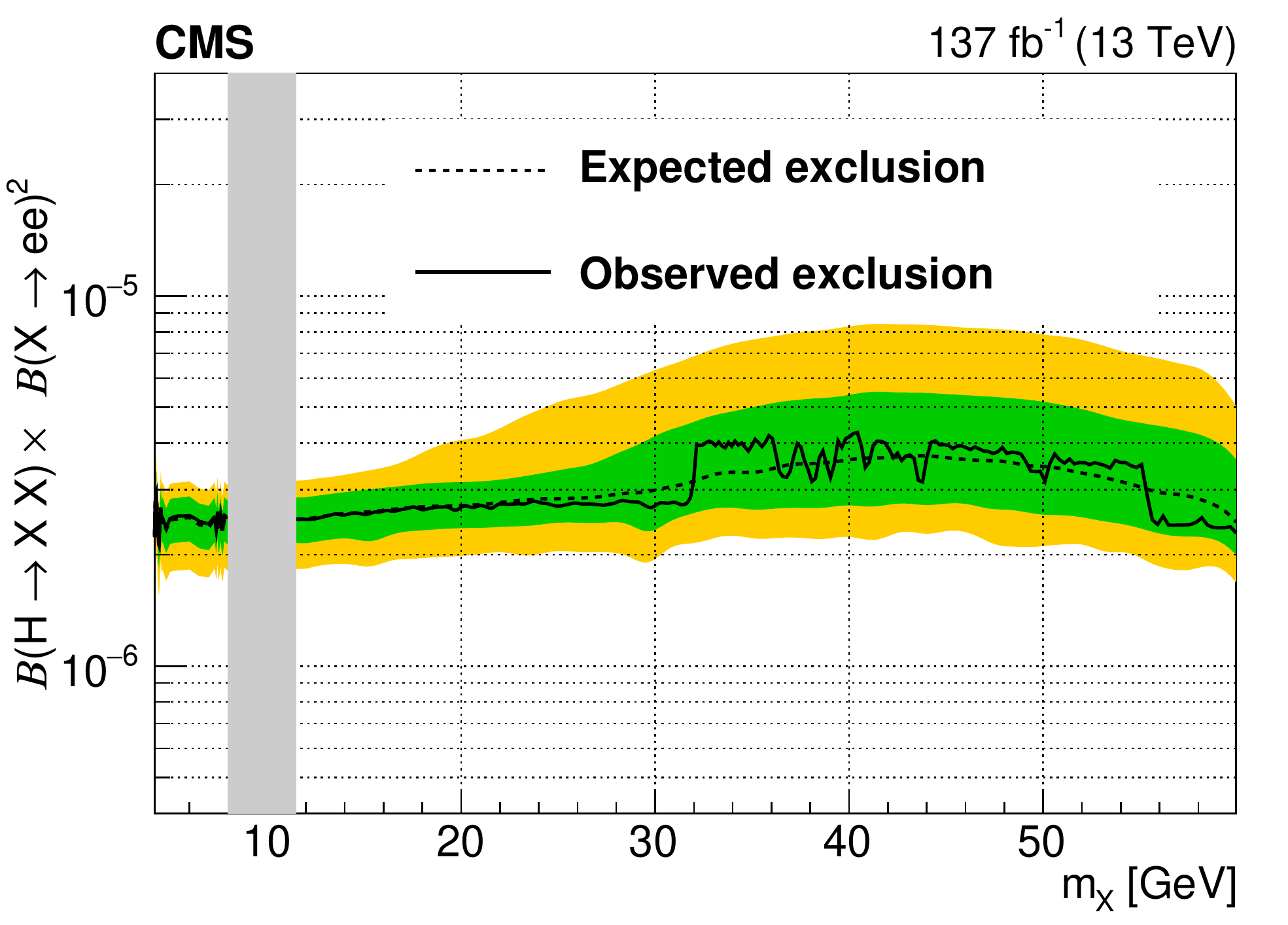}
\includegraphics[width=\cmsFigWidth]{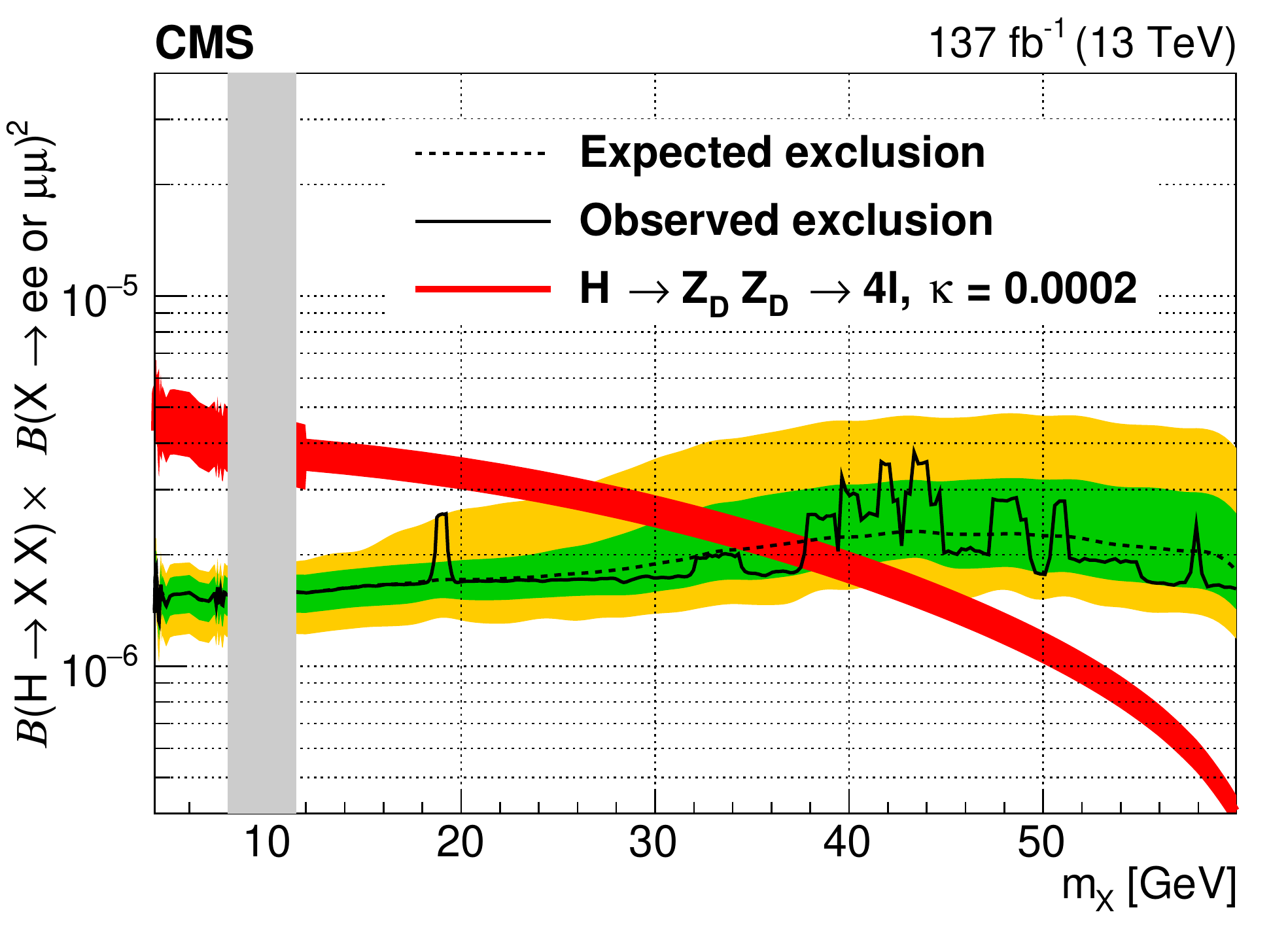}
\caption{
    Expected and observed 95\% CL limits on 
    $\mathcal{B}(\PH \to \PX \PX) \mathcal{B}(\PX \to \PGm\PGm)^2$ assuming \PX decays to dimuons only, 
    $\mathcal{B}(\PH \to \PX \PX)  \mathcal{B}(\PX \to \Pe\Pe)^2$ assuming \PX decays to dielectrons only,
    and $\mathcal{B}(\PH \to \PX \PX)  \mathcal{B}(\PX \to \Pe\Pe\ \text{or}\ \mu\mu)^2$ assuming a flavor symmetric decay of \PX to dimuons and dielectrons.
    The dashed black curve is the expected upper limit, with one and two standard-deviation bands shown in green and yellow, respectively.
    The solid black curve is the observed upper limit. 
    The red curve represents the theoretical cross section for the signal process $\PH \to \PX \PX \to 4\ell$. 
    The discontinuity at 12\GeV in uncertainty is due to 
    the switch from experimental to theoretical uncertainty estimates of $\mathcal{B}(\zd \to \Pe\Pe\ \text{or}\ \PGm\PGm)$, as described in Ref.~\cite{Curtin:2014cca}.
    The symbol $\kappa$ is the Higgs-mixing parameter. 
    The grey band corresponds to the excluded region around the $\PQb\PAQb$ bound states of $\Upsilon$.
    \label{fig:limit_Br_XX}
    }
\end{figure}

Upper limits at 95\% confidence level (CL) are derived on model-independent
branching fractions with the \ZX and \XX selections assuming three  
decay channels: a flavor symmetric decay of \PX to a muon or
an electron pair, exclusive \PX decays to a muon pair, and exclusive
\PX decays to an electron 
pair. Acceptance effects arising from different signal models are
included as systematic uncertainties in the signal yields 
after event selection. Little model dependence is expected as the
event selection is defined without using angular information between the leptons. 
Figures~\ref{fig:limit_Br_ZX} and \ref{fig:limit_Br_XX} show the
exclusion limits on the model-independent branching fractions with the \ZX 
and \XX selections, respectively. The weaker observed limit in the XX selection at ${\mass{\PX} \approx 18\GeV}$ is due to 
one observed data event and does not represent a significant statistical
deviation from the background hypothesis. Kinematic differences between the dark photon and ALP models are included as systematic uncertainties, as detailed in Section~\ref{sec:sys}.

\subsection{Limits on dark photon model parameters}
\begin{figure}[htb!p]
\centering
\includegraphics[width=0.48\textwidth]{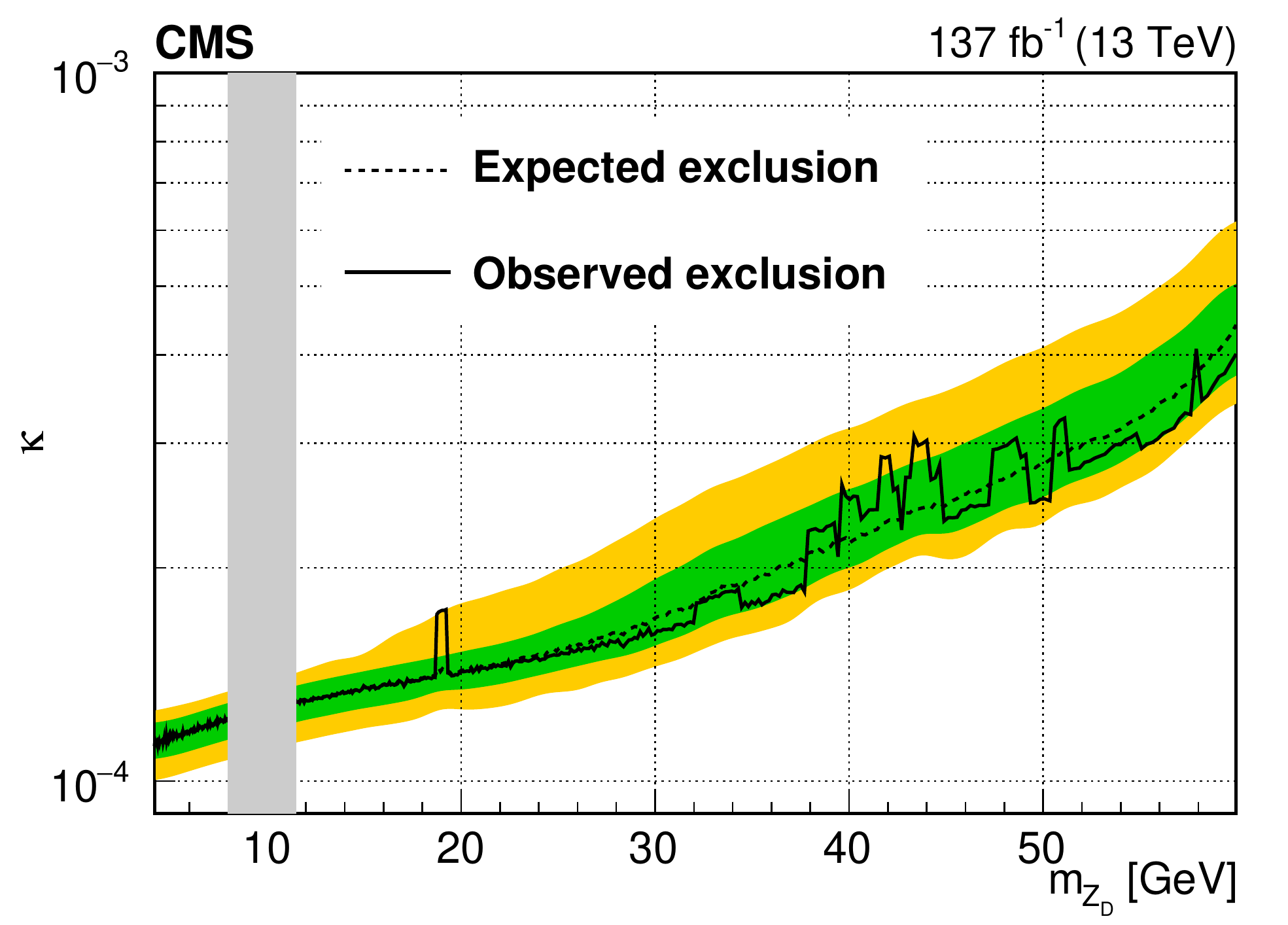}
\caption{
    95\% CL limits on the Higgs-mixing parameter $\kappa$, based on the \XX selection, as function of \mass{\zd}. The dashed black curve is the 
    expected upper limit, with one and two standard-deviation bands shown in green and yellow, respectively.
    The solid black curve is the observed upper limit.
    The grey band corresponds to the excluded region around the $\PQb\PAQb$ bound states of $\Upsilon$.
    \label{fig:limit_kappa}
    }
\end{figure}

Upper limits at 95\% CL are obtained on the Higgs-mixing parameter $\kappa$ and ${\mathcal{B}(\PH  \to \zd \zd)}$ with the \XX selection, 
as shown in Fig.~\ref{fig:limit_kappa}, assuming $\kappa \gg \esp$. 
The LHC provides unique sensitivity to the parameter $\kappa$ due to the presence 
of the Higgs boson. In addition, this analysis provides some sensitivity to $\varepsilon$, but the upper limits are almost an
order of magnitude weaker than those from the Drell--Yan search and from the LHCb Collaboration \cite{Aaij:2017rft}, and hence are not reported in this paper.

\subsection{Limits on the ALP model}
\begin{figure}[htb!]
\centering
\includegraphics[width=0.48\textwidth]{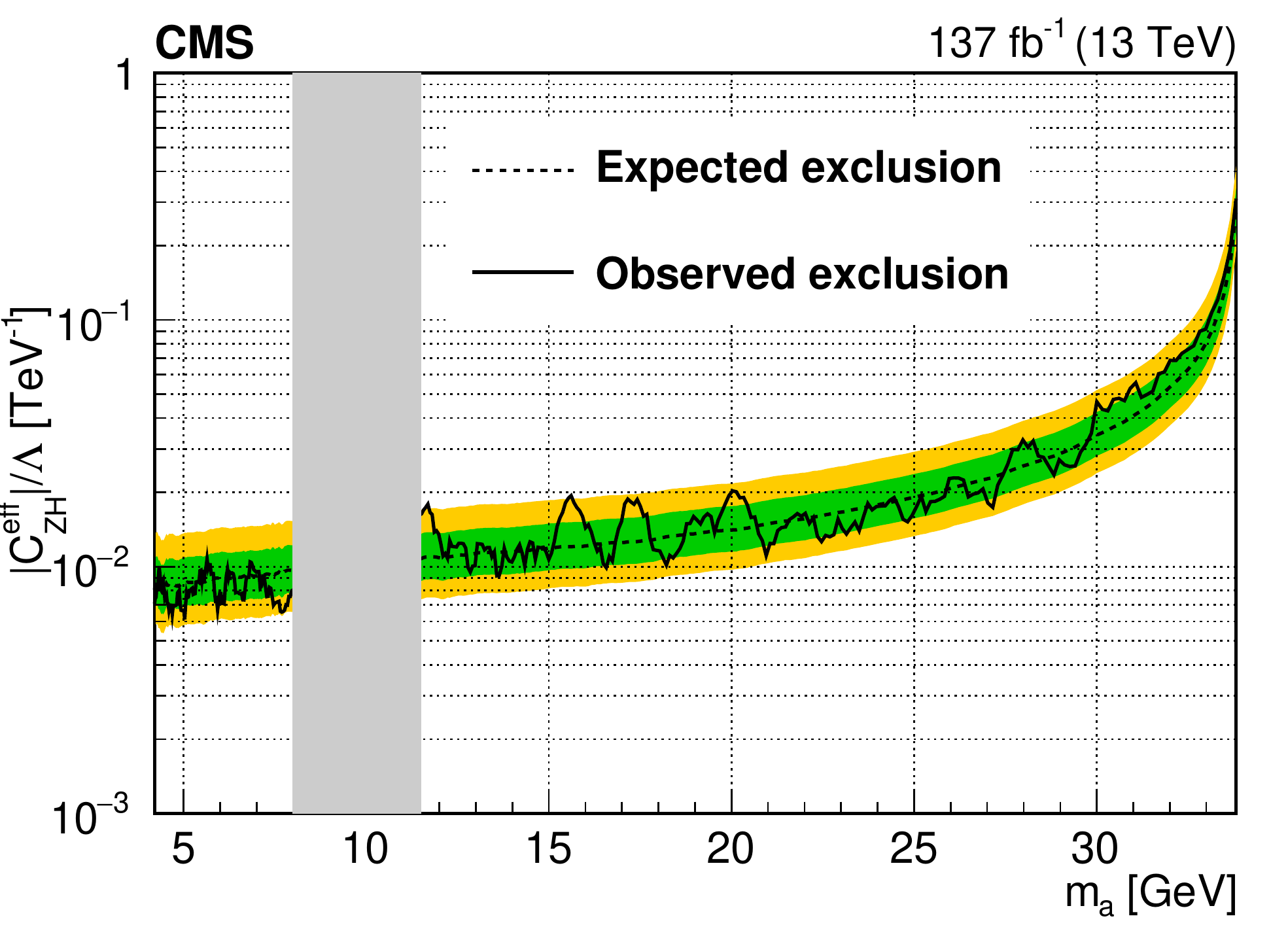}
\includegraphics[width=0.48\textwidth]{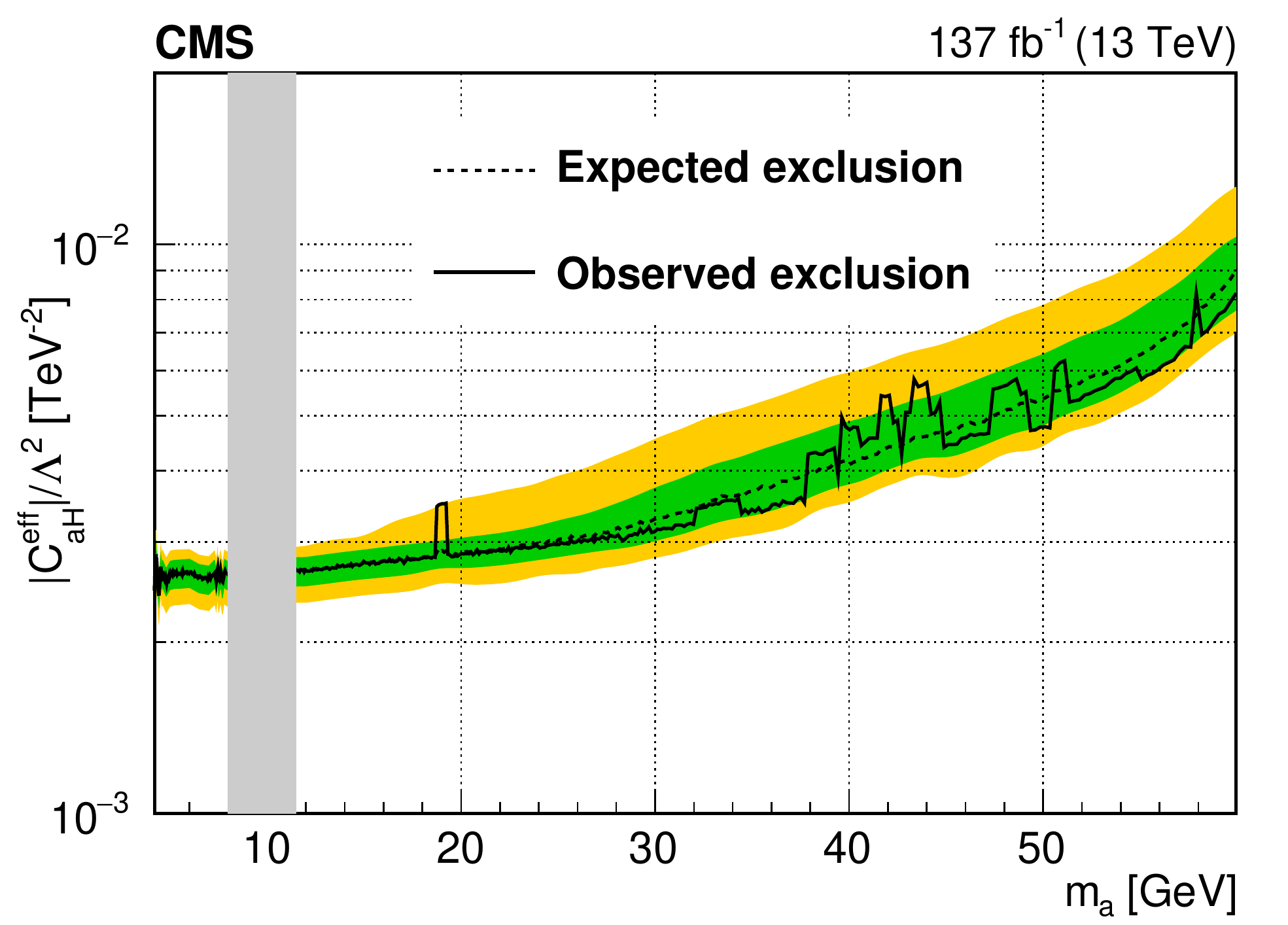}
\caption{
    95\% CL limit on \cZhLambda and \cahLambda as function of $m_{\Pa}$. Black curves are 
    the expected upper limits, with one and two standard-deviation bands shown in green and yellow, respectively.
    The solid black curves represent the observed upper limits.
    The grey band corresponds to the excluded region around the $\PQb\PAQb$ bound states of $\Upsilon$.
    \label{fig:limit_ALP}
    }
\end{figure}

Upper limits at 95\% CL are calculated on the Wilson coefficients \cZhLambda and \cahLambda, 
as shown in Fig.~\ref{fig:limit_ALP}, where \cZh is the effective coupling parameter of the Higgs boson, \PZ boson,
and the ALP, \cah is the effective coupling parameter of the Higgs boson
and the ALP, and $\Lambda$ is the new physics scale. In both interpretations, the ALP is assumed to decay promptly 
with $\mathcal{B}(\Pa\to \Pe\Pe\ \text{or}\ \PGm\PGm) = 1$, with equal fractions to muons and electrons. The last six 
mass hypotheses are omitted in the calculation of upper limits on \cZhLambda to match the \mass{\Pa} range adopted in Ref.~\cite{Bauer:2017ris}.
Kinematic differences between the dark photon and ALP models are included as corrections on signal region yields, as detailed in Section~\ref{sec:sys}.

\section{Summary}
\label{sec:summary}

A search for dilepton resonances in Higgs boson decays to four-lepton final states 
has been presented. The search considers the two intermediate decay topologies $\PH \to \PZ \PX$ and $\PH \to \PX \PX$.
No significant deviations from the standard model expectations are observed. 
The search imposes experimental constraints on products of model-independent branching fractions of
$\mathcal{B}(\PH \to \PZ \PX)$, $\mathcal{B}(\PH \to \PX \PX)$ and $\mathcal{B}(\PX \to \Pe\Pe\ \text{or}\ \PGm\PGm)$, assuming flavor-symmetric decays 
of \PX to dimuons and dielectrons, exclusive decays of \PX to dimuons, and exclusive decays of \PX to dielectrons, for $\mass{\PX} > 4\GeV$. In addition, two well-motivated theoretical frameworks beyond the standard model are considered.
Due to the presence of the Higgs boson production in LHC proton-proton collisions, 
the search provides unique constraints on the Higgs-mixing parameter $\kappa < 4 \times 10^{-4}$ at 95\% confidence level (\CL) in a dark photon model with the \XX 
selection, in Higgs-mixing-dominated scenarios, while searches for \zd in Drell--Yan processes~\cite{Sirunyan:2019wqq,Aaij:2017rft} provide better exclusion limits 
on $\varepsilon$ in kinetic-mixing-dominated scenarios. For the axion-like particle model, upper limits at 95\% CL are placed on two relevant 
Wilson coefficients \cZhLambda and \cahLambda. This is the first direct limit on decays of the observed Higgs boson to axion-like particles decaying to leptons.

\begin{acknowledgments}
\label{ack}
We congratulate our colleagues in the CERN accelerator departments for the excellent performance of the LHC and thank the technical and administrative staffs at CERN and at other CMS institutes for their contributions to the success of the CMS effort. In addition, we gratefully acknowledge the computing centres and personnel of the Worldwide LHC Computing Grid and other centres for delivering so effectively the computing infrastructure essential to our analyses. Finally, we acknowledge the enduring support for the construction and operation of the LHC, the CMS detector, and the supporting computing infrastructure provided by the following funding agencies: BMBWF and FWF (Austria); FNRS and FWO (Belgium); CNPq, CAPES, FAPERJ, FAPERGS, and FAPESP (Brazil); MES and BNSF (Bulgaria); CERN; CAS, MoST, and NSFC (China); MINCIENCIAS (Colombia); MSES and CSF (Croatia); RIF (Cyprus); SENESCYT (Ecuador); MoER, ERC PUT and ERDF (Estonia); Academy of Finland, MEC, and HIP (Finland); CEA and CNRS/IN2P3 (France); BMBF, DFG, and HGF (Germany); GSRI (Greece); NKFIA (Hungary); DAE and DST (India); IPM (Iran); SFI (Ireland); INFN (Italy); MSIP and NRF (Republic of Korea); MES (Latvia); LAS (Lithuania); MOE and UM (Malaysia); BUAP, CINVESTAV, CONACYT, LNS, SEP, and UASLP-FAI (Mexico); MOS (Montenegro); MBIE (New Zealand); PAEC (Pakistan); MSHE and NSC (Poland); FCT (Portugal); JINR (Dubna); MON, RosAtom, RAS, RFBR, and NRC KI (Russia); MESTD (Serbia); SEIDI, CPAN, PCTI, and FEDER (Spain); MOSTR (Sri Lanka); Swiss Funding Agencies (Switzerland); MST (Taipei); ThEPCenter, IPST, STAR, and NSTDA (Thailand); TUBITAK and TAEK (Turkey); NASU (Ukraine); STFC (United Kingdom); DOE and NSF (USA).

\hyphenation{Rachada-pisek} Individuals have received support from the Marie-Curie programme and the European Research Council and Horizon 2020 Grant, contract Nos.\ 675440, 724704, 752730, 758316, 765710, 824093, 884104, and COST Action CA16108 (European Union); the Leventis Foundation; the Alfred P.\ Sloan Foundation; the Alexander von Humboldt Foundation; the Belgian Federal Science Policy Office; the Fonds pour la Formation \`a la Recherche dans l'Industrie et dans l'Agriculture (FRIA-Belgium); the Agentschap voor Innovatie door Wetenschap en Technologie (IWT-Belgium); the F.R.S.-FNRS and FWO (Belgium) under the ``Excellence of Science -- EOS" -- be.h project n.\ 30820817; the Beijing Municipal Science \& Technology Commission, No. Z191100007219010; the Ministry of Education, Youth and Sports (MEYS) of the Czech Republic; the Deutsche Forschungsgemeinschaft (DFG), under Germany's Excellence Strategy -- EXC 2121 ``Quantum Universe" -- 390833306, and under project number 400140256 - GRK2497; the Lend\"ulet (``Momentum") Programme and the J\'anos Bolyai Research Scholarship of the Hungarian Academy of Sciences, the New National Excellence Program \'UNKP, the NKFIA research grants 123842, 123959, 124845, 124850, 125105, 128713, 128786, and 129058 (Hungary); the Council of Science and Industrial Research, India; the Latvian Council of Science; the Ministry of Science and Higher Education and the National Science Center, contracts Opus 2014/15/B/ST2/03998 and 2015/19/B/ST2/02861 (Poland); the Funda\c{c}\~ao para a Ci\^encia e a Tecnologia, grant CEECIND/01334/2018 (Portugal); the National Priorities Research Program by Qatar National Research Fund; the Ministry of Science and Higher Education, projects no. 14.W03.31.0026 and no. FSWW-2020-0008, and the Russian Foundation for Basic Research, project No.19-42-703014 (Russia); the Programa Estatal de Fomento de la Investigaci{\'o}n Cient{\'i}fica y T{\'e}cnica de Excelencia Mar\'{\i}a de Maeztu, grant MDM-2015-0509 and the Programa Severo Ochoa del Principado de Asturias; the Stavros Niarchos Foundation (Greece); the Rachadapisek Sompot Fund for Postdoctoral Fellowship, Chulalongkorn University and the Chulalongkorn Academic into Its 2nd Century Project Advancement Project (Thailand); the Kavli Foundation; the Nvidia Corporation; the SuperMicro Corporation; the Welch Foundation, contract C-1845; and the Weston Havens Foundation (USA).
\end{acknowledgments}

\bibliography{auto_generated}
\cleardoublepage \appendix\section{The CMS Collaboration \label{app:collab}}\begin{sloppypar}\hyphenpenalty=5000\widowpenalty=500\clubpenalty=5000\vskip\cmsinstskip
\textbf{Yerevan~Physics~Institute, Yerevan, Armenia}\\*[0pt]
A.~Tumasyan
\vskip\cmsinstskip
\textbf{Institut~f\"{u}r~Hochenergiephysik, Vienna, Austria}\\*[0pt]
W.~Adam, T.~Bergauer, M.~Dragicevic, J.~Er\"{o}, A.~Escalante~Del~Valle, R.~Fr\"{u}hwirth\cmsAuthorMark{1}, M.~Jeitler\cmsAuthorMark{1}, N.~Krammer, L.~Lechner, D.~Liko, T.~Madlener, I.~Mikulec, F.M.~Pitters, N.~Rad, J.~Schieck\cmsAuthorMark{1}, R.~Sch\"{o}fbeck, M.~Spanring, S.~Templ, W.~Waltenberger, C.-E.~Wulz\cmsAuthorMark{1}, M.~Zarucki
% ...
\vskip\cmsinstskip
% ...
1:  Also at TU~Wien, Wien, Austria\\
% ...
\end{sloppypar}
\end{document}